\documentclass[preprint,number,12pt]{elsarticle}  

\usepackage{amssymb}
\usepackage{amsmath,amssymb,amsfonts}
\usepackage{amsthm}
\usepackage{dsfont}
\usepackage{graphicx}
\usepackage{subfigure}
\usepackage{amsbsy}
\usepackage{mathrsfs}
\usepackage{lipsum}
\usepackage{color}
\usepackage{hyperref} 

\journal{European Jornal of Control}

\usepackage{comment}
\usepackage{siunitx}

\newcommand{\MFa}[1]{\textcolor{black}{#1}}

\newcommand{\rev}[1]{\textcolor{black}{#1}}

\newtheorem{Def}{\textbf{Definition}}

\newtheorem{theorem}{\textbf{Theorem}}
\newtheorem{prop}{\textbf{Proposition}}
\newtheorem{rmk}{\textbf{Remark}}
\newtheorem{corollary}{\textbf{Corollary}}

\newcommand{\drm}{\mathrm{d}}
\newcommand{\ee}{\mathbf{e}}
\newcommand{\ff}{\mathbf{f}}

\newcommand{\ww}{\mathbf{w}}

\newcommand{\psps}{\boldsymbol{\psi}}
\newcommand{\uups}{\boldsymbol{\upsilon}}
\newcommand{\lamlam}{\boldsymbol{\lambda}}

\newcommand{\xx}{\mathbf{x}}

\newcommand{\pp}{\mathbf{p}}

\newcommand{\uu}{\mathbf{u}}

\newcommand{\yy}{\mathbf{y}}

\newcommand{\Bb}{\mathbf{B}}
\newcommand{\Cc}{\mathbf{C}}

\newcommand{\Ii}{\mathbf{I}}

\newcommand{\Ll}{\mathbf{L}}
\newcommand{\Mm}{\mathbf{M}}
\newcommand{\Nn}{\mathbf{N}}

\newcommand{\Qq}{\mathbf{Q}}
\newcommand{\Rr}{\mathbf{R}}

\newcommand{\Zz}{\mathbf{O}}
\newcommand{\Thth}{\boldsymbol{\Theta}}
\newcommand{\omom}{\boldsymbol{\omega}}
\newcommand{\Omom}{\boldsymbol{\Omega}}

\newcommand{\vc}[1]{\mathrm{vec}\left({#1}\right)}
\newcommand{\vci}[2]{\mathrm{vec}_{i=1}^{#2}({#1})}
\newcommand{\vcj}[2]{\mathrm{vec}_{j=1}^{#2}({#1})}

\newcommand{\eto}[1]{\mathrm{e}^{#1}}

\renewcommand{\imath}{\boldsymbol{\mathrm{i}}}

\newcommand{\rk}{\mathrm{rk}}

\newcommand{\Diag}{\mathrm{Diag}}

\newcommand{\rig}{\boldsymbol{\mathcal{R}}}

\begin{document}

\begin{frontmatter}

\title{Optimal Time-Invariant Distributed Formation Tracking for Second-Order Multi-Agent Systems} 

\author{Marco Fabris\fnref{myfootnote}, Giulio Fattore and Angelo Cenedese}
\address{Department of Information Engineering, University of Padova, via Gradenigo 6/B, Padova, 35131, Italy.}
\fntext[myfootnote]{M.~Fabris, G.~Fattore and A.~Cenedese are with the Department of Information Engineering at the University of Padova (UNIPD - DEI), Padova, 35131, Italy (e-mails: marco.fabris.1@unipd.it, giulio.fattore@phd.unipd.it, angelo.cenedese@unipd.it). A.~Cenedese is also with the Institute of Electronics, Information and Telecommunication Engineering, National Research Council (CNR - IEIIT).}

%
%
	
\begin{abstract}
This paper addresses the optimal time-invariant formation tracking problem with the aim of providing a distributed solution for multi-agent systems with second-order integrator dynamics. In the literature, most of the results related to multi-agent formation tracking do not consider energy issues while investigating distributed feedback control laws. 
In order to account for this crucial design aspect, we contribute by formalizing and proposing a solution to an optimization problem that encapsulates trajectory tracking, distance-based formation control and input energy minimization, through a specific and key choice of potential functions in the optimization cost. 
To this end, we show how to compute the inverse dynamics in a centralized fashion by means of the Projector-Operator-based Newton's method for Trajectory Optimization (PRONTO) and, more importantly,  we exploit such an offline solution as a general reference to devise a stabilizing online distributed control law. 
Finally, numerical examples involving a cubic formation following a chicane-like path in the 3D space are provided to validate the proposed control strategies.
\end{abstract}


\end{frontmatter}



\section{Introduction}\label{sec:intro}


Over the last years, distributed control of multi-agent systems (MASs) has received considerable attention in numerous and diverse scientific communities, due to its open theoretical challenges and extensive application fields~\cite{MesbahiEgerstedt2010,LiDuan2015}. 
A MAS consists of a set of autonomous agents that take decisions and communicate information thanks to their ability to sense the environment and are able to cooperate and perform complex tasks that a single individual may not be fit to complete~\cite{rizk2019cooperative}. 

When considering a multitude of robotic agents, tasks related to trajectory optimization, path planning, and formation stabilization arise, which require solution approaches that trade-off between the requirements and constraints of both individual agents and the overall ensemble. 
In this context, a general and overarching canonical problem to be addressed is that of formation control (FC)~\cite{anderson2008rigid}, which spans from the creation and maintenance of a given spatial MAS shape (also referred to as formation producing or stabilization, FS) to the coordinated motion of the MAS configuration along a trajectory (also referred to as formation tracking, FT)~\cite{ahn2020book}. 
In the extremely vast and multi-faceted literature related to FC, some common interesting aspects can be identified.
For example, the specific features of FC depend on the level of interaction within the MAS, namely if the approach is centralized~\cite{HauserCook2016}, distributed~\cite{Hu2021ADecentralized, bishop2015distributed},
or leader-follower~\cite{trinh2018bearing, Franchi2019Online}, and on the control and sensing capabilities of the agents, basically distinguishing among position/distance information~\cite{OhParkAhn2015, ChuangHuangDOrsogna2007}, orientation/bearing information~\cite{eren2012formation, michieletto2021unified} ~\cite{zili2022ACCbearing}, or both~\cite{bishop2015distributed}.
Moreover, in order to implement these procedures and paradigms, 
FC can exploit flexible optimization approaches \cite{yan2018optimally, Hu2021ADecentralized, yan2022optimal} and make use of potential-based functions~\cite{ChuangHuangDOrsogna2007,  HauserCook2016} or alternatively, it resorts to formal geometrical approaches as those related to rigidity theory~\cite{anderson2008rigid, michieletto2021unified}\cite{xu2020affine}.

Delving more specifically into the FT problem, it may be observed how leader-follower schemes provide a popular approach to make a multiagent formation follow a desired trajectory while preserving inter-distances or bearing measurements, with the constraints imposed by the specific dynamics.  
For example, in~\cite{SalinasBricarie2017} a control strategy for ground robots with first-order dynamics is devised, where a leader agent is aware of the desired position and velocity of the formation and hence it is tasked with trajectory tracking, while the other (follower) agents arrange collision-free formation creation and maintenance. 
Leader and follower agents are also considered in~\cite{zhao2019bearing} with reference to bearing-only measurements and different classes of MASs, and in~\cite{Minh2021RobustTC} the approach is extended to also account for measurement uncertainties.
Along this line, while keeping the same control strategy, specific attention on the sensing part is given in~\cite{hu2010distributed}, where a distributed estimation solution is proposed to mitigate the presence of noisy measurements, and in\cite{Duan2020FinitetimeTO}, where the attention is placed on the followers estimation capability in finite-time, assuming heterogeneous linear systems. 
Also, in~\cite{YangCaoGarcia2018}, it is presented a distributed approach to FT that leverages an estimation strategy to obtain information on a virtual leader representing the formation centroid. 
An in-principle similar solution is discussed in~\cite{brinon2014cooperative}, where it is decoupled the problem of FS from that of FT: the former is obtained through consensus, while the latter is pursued through barycenter tracking. 
A leader-follower strategy is also employed in \cite{liu2014finite}, in which the authors exploit optimal control techniques to develop a trajectory tracking with a rigid formation, limiting, though, their analysis to the planar case with first-order dynamics.

The research in \cite{ZhangZhouWang2022} examines the issue of bipartite time-varying output formation tracking in heterogeneous MASs with multiple leaders and switching communication networks, taking into account the possibility of connected or disconnected networks. To tackle this challenge, a new approach is introduced, which involves a reduced-dimensional observer-based fully distributed asynchronous dynamic edge-event-triggered output feedback control protocol. 
With respect to fully distributed control, the approaches based on a physical leader-follower structure require either an additional procedure for the leader election~\cite{Franchi2019Online} or an a-priori or online selection based on the MAS features~\cite{Dong2017Time-Varying}.
%
	Remarkably, the formulation of an optimal problem that refers to the MAS properties and control variables allows combining both the performance aspects related to the FT task and the more practical aspects brought in by the physical features of the MAS of interest, thus leading to efficiency and feasibility of the control solution.
	For example, the geometry of the formation is specifically addressed in~\cite{LiuZhaoChen2016} and~\cite{PengSunGeng2019} to guarantee the consistency of a MAS whose agent states and behavior dynamics are defined in the special Euclidean group SE(3).
	On the other hand, the practical actuation requirements of a robotic cooperative system are considered with reference to a group of underactuated autonomous underwater vehicles in~\cite{LiXieYan2017}, where an optimization approach based on the so-called receding horizon algorithm is exploited to accommodate saturation issues. 
	However, in the latter works, as well as in the majority of the surveyed literature, 
    the simultaneous optimization of formation, tracking, and energy consumption is generally overlooked,
    even though this may represent a crucial design principle impacting both the task completion effectiveness and the control action efficiency.

\textit{Contributions}: Given this premise, in this paper we propose a novel approach to FT in which the problem is formulated based on a time-invariant optimal control objective where three different tasks are considered, namely that of tracking a given path at the group level, attaining a desired (and feasible) geometric formation while minimizing the input energy. Compared to the work in \cite{yu2022adaptive}, the approach we pursue is fully distributed and attains finite time optimal control.
Also, differently from \cite{Dong2017Time-Varying, dong2019TVswitching} in which the network topology is a switching one, 
here we consider a more realistic scenario by allowing for the presence of communication constraints in the system of mobile robots, whenever the information exchange between a pair of agents is not ensured. Moreover, unlike \cite{loria2016leaderfollower,dong2017TVswitching} in which one or more leader agent are identified, our work is developed assuming the absence of hierarchy among the agents.
More specifically, we build on our previous work~\cite{FabrisCenedeseHauser2019}, where the optimization framework called PRONTO (PRojection Operator based Newton's method for Trajectory Optimization) was successfully adopted to obtain inverse dynamics to steer a MAS towards the desired formation along the chosen path. \rev{Then, following the Pontryagin's Maximum Principle (PMP) \cite{Pontryagin2018},} we devise in the present work an online networked strategy
resting on a \rev{PMP}-based distributed feedback control law.
%

\textit{Outline}: The remainder of the paper is organized as follows. 
Sec.~\ref{sec:problem_setup} deals with the general setup, providing the basic notation and describing the adopted multi-agent models and dynamics.
The first part of Sec.~\ref{sec:numerical_methodologies_for_OIFT} introduces, in a general way, how the FT problem of interest can be formulated and optimally solved in a centralized fashion by using PRONTO; whereas, in the second part, the derivation of a stabilizing online distributed feedback controller is presented. 
Lastly, Sec.~\ref{sec:numerical_simulations} illustrates some numerical results, providing a validation of the proposed approach and interesting remarks, and Sec.~\ref{sec:conclusions} sketches out future directions for this research.

\section{General setup}\label{sec:problem_setup}
In this section, the models and assumptions are provided to formalize the optimal time-invariant formation tracking (OIFT) problem, enriching the setup introduced in~\cite{FabrisCenedeseHauser2019}. In addition, communication topology constraints are taken into account for the development of a related distributed control law.


\subsection{Multi-agent model and basic notation}
The multi-agent system under analysis is topologically represented by an undirected graph $\mathcal{G} = (\mathcal{V},\mathcal{E})$ formed by the set of nodes $\mathcal{V}$ and the set of edges $\mathcal{E}$. 
Each node of $\mathcal{V}$ is simply addressed with index $i = 1,\ldots,n$, where $n = |\mathcal{V}|$ is the cardinality of $\mathcal{V}$ and node $i$ is referred to as the $i$-th agent. Also, node $i$ is allowed to exchange information with node $j$ if and only if there exists an edge $(i,j)=(j,i)$ contained in $\mathcal{E}$. 
We assume that $\mathcal{G}$ is connected, i.e. there exists a collection of edges that link each node $i$ to any other node $j \neq i$. The neighborhood of node $i$ is defined as the set $\mathcal{N}_{i} = \{j\in \mathcal{V}|(i,j)\in \mathcal{E}\}$. 
%
Lastly, the Laplacian matrix associated to $\mathcal{G}$ is defined as $\mathbf{L}=\mathbf{D}-\mathbf{A}$, where $\mathbf{A}\in \mathbb{R}^{n\times n}$ denotes the adjacency matrix of $\mathcal{G}$, such that $[\mathbf{A}]_{ij} = 1$ if $(i,j) \in \mathcal{E}$; $[\mathbf{A}]_{ij} = 0$ otherwise, and $\mathbf{D}\in \mathbb{R}^{n\times n}$ denotes the degree matrix of $\mathcal{G}$, such that $\mathbf{D}$ is diagonal and $[\mathbf{D}]_{ii} = \sum_{j=1}^{n} [\mathbf{A}]_{ij}$.
The remaining basic notation follows.

The set of real numbers is indicated by $\mathbb{R}$ and a similar symbology is straightforwardly employed for further subsets or extensions of $\mathbb{R}$ itself. The class of $k$-continuously differentiable 
functions is identified with $\mathscr{C}^{k}$.

The identity matrix of dimension $\natural \in \mathbb{N}$ is denoted by $\Ii_{\natural}$, while $\Zz_{\natural_{1}\times \natural_{2}}$ addresses the zero matrix with dimensions $\natural_{1}\times \natural_{2} \in \mathbb{N} \times \mathbb{N}$. Moreover, if $\natural_{1}= \natural_{2}=\natural$ symbol $\Zz_{\natural}$ is used instead of $\Zz_{\natural_{1}\times \natural_{2}}$, while the zero vector of dimension $\natural$ is denoted by $\mathbf{0}_{\natural}$. The rank of a matrix is expressed by the operator $\rk$. Symbols $\succeq 0$, $\succ 0$, and $\left\|\cdot \right\|_{2}$ stand for positive semidefinite, positive definite, and Euclidean norm, respectively. Whereas, $\left\|\psps\right\|_{\Omom}^{2} $ defines the quadratic form $\psps^{\top}\Omom \psps$, where $\psps$ and $\Omom \succ 0$ generically indicate real-valued vector and a real-valued matrix, respectively.
Two kinds of vectorization operators are defined: $\mathrm{vec}$ and $\mathrm{vec}_{i=1}^{\natural}$. The former either takes all its vector arguments $\psps_{1} ,\dots, \psps_{\natural}$ and stack them into a single vector $ \vc{\psps_{1} ,\dots, \psps_{\natural}} = [\psps_{1}^{\top} ~\cdots~ \psps_{\natural}^{\top}]^{\top}$ or transforms the columns of a matrix $\Omom = [\omom_{1} ~\cdots~ \omom_{\natural}]$ into the vector $\vc{\Omom}=[\omom_{1}^{\top} ~\cdots~ \omom_{\natural}^{\top}]^{\top}$. The latter, instead, constructs the stack vector $\vci{\psps_{i}}{\natural} = [\psps_{1}^{\top} ~\cdots~ \psps_{\natural}^{\top}]^{\top}$ starting from a given vector family $\{\psps\}_{i=1}^{\natural} = \{\psps_{1} ,\dots, \psps_{\natural}\}$. The block-diagonal operator that maps the family of real matrices $\{\Omom\}_{i=1}^{\natural} = \{\Omom_{1},\dots,\Omom_{\natural}\}$ into its associated diagonal matrix is denoted with $\Omom = \mathrm{Diag}(\Omom_{1},\dots,\Omom_{\natural})$. 

Variable $t\geq 0$ specifies the continuous time instants, while symbol $(\cdot)$ denotes a trajectory as $t$ spans interval $[0,T]$, for a chosen $T>0$. Symbols $\nabla_{*}$ and $\mathcal{H}_{**}$ indicate standard gradient and Hessian operators w.r.t. some generic variable~$*$. Finally, symbol $\dot{\nabla}_{*}$ is equivalent to $\dfrac{\drm}{\drm t} \nabla_{*}$.

\subsection{Dynamics of the agents}
We suppose that $n\geq 2$ robotic agents with linear dynamics are already deployed in an environment space of dimension $M$, such that $M\in \left\lbrace 1,2,3\right\rbrace  $. We also assume that each agent $i$ is aware of its absolute position $\mathbf{p}_{i} =\mathbf{p}_{i}(t) \in \mathbb{R}^{M}$ and velocity $\dot{\mathbf{p}}_{i}\in \mathbb{R}^{M}$ and can be governed by means of a regulation on its absolute acceleration $\uu_i=\ddot{\mathbf{p}}_{i}(t)\in \mathbb{R}^{M}$. 
Setting $N=nM$, the expressions of the state $\mathbf{x}\in \mathbb{R}^{2N}$ and the input $\mathbf{u}\in \mathbb{R}^{N}$ for this group of mobile elements are provided respectively by
\begin{align*}
	\mathbf{x} &= \begin{bmatrix}
		\mathbf{p}_{1}^{\top} & \dots & \mathbf{p}_{n}^{\top} & \dot{\mathbf{p}}_{1}^{\top} & \dots & \dot{\mathbf{p}}_{n}^{\top} 
	\end{bmatrix}^{\top} = \begin{bmatrix}
		\mathbf{p}^{\top} & \dot{\mathbf{p}}^{\top}
	\end{bmatrix}^{\top};\\
	\mathbf{u} &= 
	\begin{bmatrix}
        \ddot{\mathbf{p}}_{1}^{\top} & \dots & \ddot{\mathbf{p}}_{n}^{\top}
	\end{bmatrix}^{\top} = \ddot{\mathbf{p}}.
\end{align*}
Equivalently, exploiting the vectorization operators, one has $\pp = \vci{\pp_{i}}{n}$, $\dot{\pp} = \vci{\dot\pp_{i}}{n}$, $\xx = \vc{\pp,\dot{\pp}}$ and $\uu = \vci{\uu_{i}}{n}$.

Differently from \cite{FabrisCenedeseHauser2019}, we assume that the state information is not globally accessible for each agent in the online control law to be provided so that only a local estimate of the centroid position $\mathbf{p}_{c} =  n^{-1} \sum_{i=1}^{n} \mathbf{p}_{i}$ and velocity $\dot{\mathbf{p}}_{c}$ is available to each agent $i$ at each time $t$. For this reason, the undirected graph $\mathcal{G}$ representing the communication network among the agents is supposed to be connected. Also, we set
$\mathbf{x}_{c} = \vc{\mathbf{p}_{c},\dot{\mathbf{p}}_{c}}$,
	with $\mathbf{x}_{c}\in \mathbb{R}^{2M}$. 
	Since we desire to steer the agents by controlling their positions and velocities through their accelerations, the classic double integrator model is chosen. 
	The latter model is described by the following linear state-space equations with second-order integrator dynamics 
	\begin{equation}\label{eq:dyn_sys}
		\begin{cases}
			\dot{\mathbf{x}}(t) = \mathbf{A} \mathbf{x}(t) + \mathbf{B} \mathbf{u}(t)\\
			\xx(0) = \mathbf{x}_{0} \in \mathbb{R}^{2N}
		\end{cases}, \qquad \xx_{c}= \mathbf{C}\mathbf{x};
	\end{equation}
	where the state matrix $\mathbf{A}\in \mathbb{R}^{2N\times 2N}$, the input matrix $\mathbf{B}\in \mathbb{R}^{2N\times N}$ and the centroid matrix $\mathbf{C} \in \mathbb{R}^{2M\times 2N}$ take the form  
	%
			%
			\begin{equation*}
					\mathbf{A}=\begin{bmatrix}
						\Zz_{N} & \mathbf{I}_{N}\\
						\Zz_{N} & \Zz_{N}
					\end{bmatrix}, 
					\quad
					\mathbf{B} = \begin{bmatrix}
						\Zz_{N}\\
						\mathbf{I}_{N}
					\end{bmatrix},
					\quad
					\mathbf{C} = \dfrac{1}{n} \begin{bmatrix}
						\mathbf{I}_{M} & \cdots & \mathbf{I}_{M} & \Zz_{M} & \cdots & \Zz_{M}\\
						\Zz_{M}  & \cdots &  \Zz_{M} & \mathbf{I}_{M} & \cdots & \mathbf{I}_{M} 
					\end{bmatrix}.
			\end{equation*}
			%
			%
			%
			

\subsection{Problem formulation} \label{subsec:problemformul}
The purpose of this study is the design and analysis of control strategies for the group of agents to track a desired path 
$\mathbf{x}_{c,des}(\cdot)= \vc{\mathbf{p}_{c,des}(\cdot),\dot{\mathbf{p}}_{c,des}(\cdot)}$ 
with its centroid $\mathbf{x}_{c}(\cdot)$, while minimizing the energy spent by the input $\mathbf{u}(\cdot)$ and attaining a desired distance-based formation. 
To this aim, we formalize such requirements as a finite-time optimal control problem as follows and, in the sequel, we devise a (practically) optimal distributed feedback controller.

Let $\mathscr{T}$ be the trajectory manifold of \eqref{eq:dyn_sys}, such that $\boldsymbol{\xi} = \left(\mathbf{x}(\cdot),\mathbf{u}(\cdot)\right) \in \mathscr{T}$ denotes a state-input trajectory. The general OIFT problem can be stated as follows: find a solution $\boldsymbol{\xi}^{\star}$ such that
\begin{equation}\label{eq:minimizationproblemOIFT}
	\underset{\boldsymbol{\xi}\in \mathscr{T}}{\min}~ h(\boldsymbol{\xi})
\end{equation}
is attained, where
\begin{equation}\label{eq:cost}
	h(\mathbf{x}(\cdot),\mathbf{u}(\cdot)) = m(\mathbf{x}(T)) + \int_{0}^{T} l\left(\mathbf{x}(\tau), \mathbf{u}(\tau) \right) \drm \tau 
\end{equation}
represents the cost functional to be minimized over the time interval $[0,T]$ in order to achieve the three objectives previously introduced. Generalizing the approach in \cite{FabrisCenedeseHauser2019}, two different terms explicitly appear in \eqref{eq:cost}:
the instantaneous cost
\begin{equation}\label{eq:inst_cost}
	l\left(\mathbf{x}(t), \mathbf{u}(t) \right) = l^{tr}(\mathbf{x}_{c}(t)) + l^{in}(\mathbf{u}(t)) + l^{fo1}(\mathbf{p}(t)) + l^{fo2}(\dot{\mathbf{p}}(t)) 
\end{equation}
and the final cost
\begin{equation}\label{eq:final_cost_m}
	m(\mathbf{x}(T)) = l^{tr}(\mathbf{x}_{c}(T)) + l^{fo1}(\mathbf{p}(T)) + l^{fo2}(\dot{\mathbf{p}}(T)) .
\end{equation}
Each term in \eqref{eq:inst_cost} is minimized with the purpose to obtain the fulfillment of a peculiar task. Indeed, setting 
$\ee_{ij}(t) = \mathbf{p}_{i}(t)-\mathbf{p}_{j}(t)$ and $s_{ij}(t) = \left\| \ee_{ij}(t) \right\|_{2}^{2}$
as the displacement of agent $j$ w.r.t. to agent $i$ and the $(i,j)$-th squared distance, respectively, each contribution can be defined as:
\begin{equation}\label{eq:track_inst_cost}
	l^{tr}(\mathbf{x}_{c}(t)) = \dfrac{1}{2} \sum\limits_{i=1}^{n} \left\|\mathbf{x}_{c}(t)-\mathbf{x}_{c,des}(t)\right\|_{\mathbf{Q}_{i}}^{2} 
\end{equation}
for the tracking task,
\begin{equation}\label{eq:inst_energy}
	l^{in}(\mathbf{u}(t)) = \dfrac{1}{2} \sum\limits_{i=1}^{n} \left\|\mathbf{u}_{i}(t)\right\|_{\mathbf{R}_{i}}^{2} 
\end{equation}
for the input energy task and, given a family of potential functions $\sigma_{d_{ij}}:\mathbb{R}_{\geq 0} \rightarrow \mathbb{R}_{\geq 0}$, such that $(i,j) \in \mathcal{E}$,
\begin{equation}\label{eq:form_inst_cost}
	l^{fo1}(\mathbf{p}(t)) = \dfrac{k_{F}}{4} \sum\limits_{i=1}^{n} \sum\limits_{j \in \mathcal{N}_{i}} \sigma_{d_{ij}}\left(s_{ij}(t)\right) 
\end{equation}
for the formation task. Some slight modifications w.r.t. what designed in \cite{FabrisCenedeseHauser2019}  have been accomplished to normalize and make the overall functional \eqref{eq:cost} distributable. Remarkably, this does not affect conceptually the original version of the OIFT problem. Moreover, to fine-tune the agents' displacement rate, we also account for the formation term associated with the inter-agent velocities
\begin{equation}\label{eq:align_inst_cost}
	l^{fo2}(\dot{\mathbf{p}}(t)) = \dfrac{k_{A}}{4} \sum\limits_{i=1}^{n} \sum\limits_{j \in \mathcal{N}_{i}}  \left\| \dot{\ee}_{ij}(t) \right\|_{\Thth_{ij}}^{2} .
\end{equation}
It is worth observing that terms \eqref{eq:track_inst_cost} and \eqref{eq:form_inst_cost} lead the system to achieve second-order velocity alignment\footnote{I.e., steering towards the (prescribed) average heading of local neighbors.} and cohesion/separation\footnote{I.e., steering to achieve the prescribed inter-agent distance/ avoid crowding.}, respectively, according to the Boids model principles~\cite{Reynolds1987}. Whereas, penalty \eqref{eq:align_inst_cost} contributes mainly to achieve alignment, as it ensures stillness (up to rotations) for the attained shape w.r.t. a dynamic frame referring to the desired centroid trajectory.

Hereafter, we address the instantaneous state cost as
\begin{equation*}
	l^{st}(\mathbf{x}) = l^{tr}(\mathbf{x})+l^{fo}(\mathbf{x}),
\end{equation*}
where $l^{fo}(\mathbf{x}) = l^{fo1}(\mathbf{p})+l^{fo2}(\dot{\mathbf{p}})$. Similarly, 
we split \eqref{eq:track_inst_cost} into the additive terms $ l^{tr1}(\pp)$ and $ l^{tr2}(\dot{\pp})$ referred to the control of $\pp_{c}$ and $\dot{\pp}_{c}$, respectively,
to benefit from the structure of
$\nabla_{\mathbf{x}} l^{tr}(\xx) = \begin{bmatrix}
\nabla_{\mathbf{p}} l^{tr1}(\pp)^\top &  \nabla_{\dot{\pp}} l^{tr2}(\dot{\pp})^\top
\end{bmatrix}^\top$
 in the sequel. 
For all $i=1, \ldots, n$, we set $\mathbf{Q}_{i} = \Diag ( \mathbf{Q}_{c,i}, \mathbf{Q}_{\dot{c},i} ) \succeq 0$, such that $\mathbf{Q}_{c,i} \in \mathbb{R}^{M\times M}$, $\mathbf{Q}_{c,i} \succeq 0$ and $\mathbf{Q}_{\dot{c},i} \in \mathbb{R}^{M\times M}$, $\mathbf{Q}_{\dot{c},i} \succeq 0$ weight the $i$-th centroid position and velocity respectively, $\mathbf{R}_{i}\in \mathbb{R}^{M\times M}$, $\mathbf{R}_{i} \succ 0$, $k_{F}>0$, $k_{A}>0$ and $\Thth_{ij} \in \mathbb{R}^{M\times M}$, $\Thth_{ij} \succeq 0$, $\forall (i,j) \in \mathcal{E}$. 
All these parameters are constant, tuned according to given specifications in order to penalize the trajectory tracking error and the energy spent by the input, and to modulate the convergence to a desired shape.
Furthermore, each $d_{ij}$ in the potential functions $\sigma_{d_{ij}} $ belonging to formation cost \eqref{eq:form_inst_cost} represents the desired inter-agent distance between a pair of agents $(i,j) \in \mathcal{E}$: an accurate selection of $\sigma_{d_{ij}}$ leads to the implementation of the cohesion/separation paradigm, as explained next.

With regard to the formation objective, it is required that the system of agents achieves a desired $M$-dimensional geometric shape induced by a set of constraints of the form
\begin{equation}\label{eq:dist_constraints}
	s_{ij} = d^{2}_{ij}, \quad \forall (i,j) \in \mathcal{E}.
\end{equation}
with $d_{ij}=d_{ji}$ $\forall$ $(i,j)\in \mathcal{E}$.\\
The use of potential functions allows fulfilling this aim, 
and, among several possible choices, it has been designed the following:
\begin{equation}\label{eq:potential}
	\sigma_{d_{ij}}(s_{ij}) = \begin{cases}
		k_{r_{ij}} (1-s_{ij}/d_{ij}^{2})^{\beta_{ij}}, \quad & \text{if } 0\leq s_{ij} < d_{ij}^{2}; \\
		k_{a_{ij}} \left((s_{ij}/d_{ij}^{2})^{\alpha_{ij}}-1\right)^{\beta_{ij}}, \quad& \text{if } s_{ij} \geq d_{ij}^{2}.
	\end{cases}
\end{equation}
This is a power function in $s_{ij}$ with degree $\beta_{ij} \in \mathbb{R}$, $\beta_{ij} \geq 2$ and differentiable w.r.t. $s_{ij}\geq 0$ up to the second order. In addition, for $\alpha_{ij} \in \mathbb{R}$, $\alpha_{ij}>0$, function \eqref{eq:potential} is $\mathscr{C}^{2}$ and its magnitude can be adjusted by tuning constants $k_{r_{ij}}\geq0$ and $k_{a_{ij}}\geq0$ independently\footnote{The only exception is for $\beta_{ij}=2$, in which, to guarantee continuity of class $\mathscr{C}^{2}$, condition $k_{r_{ij}} = \alpha_{ij}^{2}k_{a_{ij}}$ must be enforced.}. 
These two parameters are purposely selected to be directly proportional to the repulsion and attraction actions between agents respectively, playing a crucial role in the intensity regulation 
of the potential itself. 
Moreover, exponent $\alpha_{ij}$ allows tuning the magnitude of the attractive potential on certain conditions. Specifically, if $s_{ij} > 2^{1/\alpha_{ij}} d_{ij}^{2}$, we say that, as $\sigma_{d_{ij}} > k_{a_{ij}}$, the attraction force becomes significantly high for the couple $(i,j)\in \mathcal{E}$. On the other hand, the maximum repulsion intensity is reached as $s_{ij}$ approaches $0$ to avoid collisions, with $\sigma_{d_{ij}} \rightarrow k_{r_{ij}}$, being $\sigma_{d_{ij}} \leq k_{r_{ij}}$ in this regime. 
A default choice may be preferred over others by setting $\alpha_{ij} = 0.5$, since this implies that the attractive potential becomes very significant whenever the effective squared inter-agent distance $s_{ij}$ exceeds the double of its reference $d_{ij}^{2}$. 
However, in practice, $\alpha_{ij}$ should be selected according to the maximum communication range allowed between $(i,j)$.

\begin{figure}[t!]
		\centering
		\includegraphics[width=0.5\textwidth]{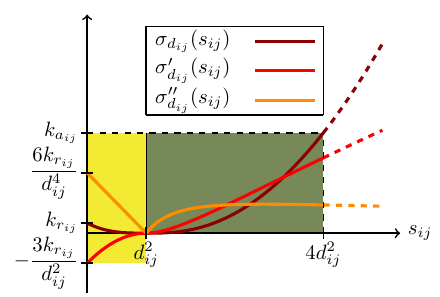}
		\caption{Potential function $\sigma_{d_{ij}}(s_{ij})$ with $\beta_{ij}=3$, $\alpha_{ij}=0.5$, $k_{a_{ij}}>k_{r_{ij}}$ and its derivatives w.r.t. $s_{ij}$ up to the second order. Attractive and repulsive behaviors can be associated to the values in the dark green and yellow areas respectively.}
		\label{fig:sigma}
	\end{figure}

The qualitative properties of functions in \eqref{eq:potential} and their 
first and second derivatives w.r.t. $s_{ij}$ (indicated with $\sigma_{d_{ij}}^{\prime}$ and $\sigma_{d_{ij}}^{\prime\prime}$ respectively) are depicted in Fig. \ref{fig:sigma} and can be summarized as follows:
\begin{subequations}
\begin{align}
     &\sigma_{d_{ij}}^{\prime} (s_{ij}) \leq 0 \ \text{for}\ 0\leq s_{ij}< d^{2}_{ij}; \label{prop:sig_prime}\\
	&\sigma_{d_{ij}}^{\prime} (s_{ij}) \geq 0 \ \text{for}\ s_{ij}\geq d^{2}_{ij};\\
	 &\sigma_{d_{ij}}^{\prime\prime} (s_{ij}) \geq 0\ \text{for all}\ s_{ij}; \label{prop:sig_prime_prime}
\end{align}
\end{subequations}
where the equalities in \eqref{prop:sig_prime}-\eqref{prop:sig_prime_prime} 
hold if and only if $s_{ij}=d^{2}_{ij}$.

	
The usage of functions with the same properties of \eqref{eq:potential} can be justified by the fact that they can lead formations to verify the maximum number of feasible relations\footnote{Meaning that such relations can be satisfied concurrently.} in \eqref{eq:dist_constraints} since the dynamics is intentionally driven to minimum-potential trajectories. Indeed, it can be easily proven that $\sigma_{d_{ij}}(s_{ij})$ is nonnegative for all $s_{ij}$ and exhibits a unique 
global minimizer $s_{ij}=d^{2}_{ij}$ for which $\sigma_{d_{ij}}(d_{ij})=0$,  
implying that \eqref{eq:form_inst_cost} vanishes if and only if the desired distance $d_{ij}$ between $(i,j)$ is achieved.


\section{Solution for the general OIFT problem} \label{sec:numerical_methodologies_for_OIFT} %
In this section, we clarify how the numerical tool PRONTO (see \cite{HauserSaccon2006,AguiarPedroBayer2017}) used for trajectory optimization is adopted to solve the OIFT problem discussed in Sec. \ref{subsec:problemformul}. 
In addition, we devise an online distributed feedback controller approximating the performance of inverse dynamics computed by means of PRONTO,  
where the latter will serve as a reference to compare the performances obtained with such a decentralized approach. The stabilizing properties ensured through such an online distributed control input are then briefly discussed together with the made approximations and heuristics to meet optimality.

\subsection{Offline centralized solution via PRONTO}\label{subsec:PRONTOsolvesOIFT}
The application of PRONTO to the OIFT problem allows to iteratively compute, yet in an offline and centralized fashion, a complete trajectory $\boldsymbol{\xi}_{k+1} = (\mathbf{x}_{k+1}(\cdot),\mathbf{u}_{k+1}(\cdot))$ satisfying\footnote{The solution exists unique as the system \eqref{eq:dyn_sys} is control affine and term $l^{in}$ is quadratic (see also \cite{HAUSER2002377} for more details).} \eqref{eq:minimizationproblemOIFT}, recalling that $\mathbf{x}_{k}(\cdot)$ and $\mathbf{u}_{k}(\cdot)$ respectively represent the absolute position and velocity state vector, and the acceleration input, for the whole evolution at the $k$-th PRONTO iteration. As mentioned previously, the assumptions made in Sec. \ref{sec:problem_setup} allow to obtain the expressions of many relevant numerical quantities. Indeed, here, we clarify how PRONTO variables and parameters 
are assigned and computed in this particular framework. Setting
$\ff(\mathbf{x},\mathbf{u}) = \mathbf{A} \mathbf{x} + \mathbf{B} \mathbf{u}$
the list of relations employed in PRONTO to find the best Newton's descent direction\footnote{For further details on the PRONTO implementation, the reader is addressed to \cite{FabrisCenedeseHauser2019}.} 
is immediately obtained and shortly reported below for convenience:  
\begin{alignat}{3}
	& \bar{\mathbf{A}}(t) &&= \ff_{\mathbf{x}} &&=
    \frac{\partial \ff(\mathbf{x},\mathbf{u})}{\partial \mathbf{x} }= \mathbf{A} \nonumber\\
	& \bar{\mathbf{B}}(t) &&= \ff_{\mathbf{u}} &&=
 \frac{\partial \ff(\mathbf{x},\mathbf{u})}{\partial \mathbf{u} }          = \mathbf{B} \nonumber\\
	& \mathbf{a}(t)       &&= l^{\top}_{\mathbf{x}}    &&= \underbrace{\mathbf{C}^{\top} \mathbf{Q} \left(\mathbf{x}_{c} -\mathbf{x}_{c,des}\right)}_{\nabla_{\xx} l^{tr}}  + \nabla_{\mathbf{x}}l^{fo1} + \nabla_{\mathbf{x}}l^{fo2} \label{eq:a}\\
	& \mathbf{b}(t)       &&= l^{\top}_{\mathbf{u}}    &&= \nabla_{\uu} l= \mathbf{R} \mathbf{u} \nonumber\\
	& \mathbf{r}_{1}(T)   &&= m^{\top}_{\mathbf{x}}    &&= \underbrace{\mathbf{C}^{\top} \mathbf{Q} \left(\mathbf{x}_{c}(T)-\mathbf{x}_{c,des}(T)\right)}_{\nabla_{\xx} l^{tr}(\xx(T))} +  \nabla_{\mathbf{x}}l^{fo1}(T) + \nabla_{\mathbf{x}}l^{fo2}(T) \nonumber\\
	& \mathbf{Q}_{o}(t)   &&= l_{\mathbf{x}\mathbf{x}} &&= \underbrace{\mathbf{C}^{\top} \mathbf{Q} \mathbf{C}}_{\mathcal{H}_{\mathbf{x}\mathbf{x}} l^{tr}} + \mathcal{H}_{\mathbf{x}\mathbf{x}} l^{fo1} + \mathcal{H}_{\mathbf{x}\mathbf{x}} l^{fo2} \label{eq:Qo}\\
	& \mathbf{S}_{o}(t)   &&= l_{\mathbf{x}\mathbf{u}} &&= \mathcal{H}_{\mathbf{x}\mathbf{u}}=\Zz_{2N\times N} \nonumber\\
	& \mathbf{R}_{o}(t)   &&= l_{\mathbf{u}\mathbf{u}} &&= \mathcal{H}_{\mathbf{u}\mathbf{u}}=\mathbf{R} \nonumber\\
	& \mathbf{P}_{1}(T)   &&= m_{\mathbf{x}\mathbf{x}} &&= \underbrace{\mathbf{C}^{\top} \mathbf{Q} \mathbf{C}}_{\mathcal{H}_{\mathbf{x}\mathbf{x}} l^{tr}(\xx(T))} + \mathcal{H}_{\mathbf{x}\mathbf{x}} l^{fo1}(T) + \mathcal{H}_{\mathbf{x}\mathbf{x}} l^{fo2}(T) \nonumber
\end{alignat}
where $\mathbf{Q} = \sum_{i=1}^{n} \mathbf{Q}_{i} $, $\mathbf{R} = \Diag ( \mathbf{R}_{1}, \ldots, \mathbf{R}_{n} ) \in \mathbb{R}^{N \times N}$.\\
In the light of this result, a time-invariant PD controller as
\begin{equation}\label{eq:PDcontroller}
	\mathbf{K} = \begin{bmatrix}
		k_{p} \mathbf{I}_{N} & k_{d} \mathbf{I}_{N}
	\end{bmatrix}, \quad k_{p}, k_{d} > 0
\end{equation}
turns out to be one of the most efficient options to adopt\footnote{The control gain \eqref{eq:PDcontroller} is a parameter used to construct the PRONTO inverse dynamics through offline state feedback. In particular, it is used to map the states into control inputs.}.

As a further observation, let us analyze the gradient and Hessian matrix of $l^{fo1}$ w.r.t. the state $\mathbf{x}$ in \eqref{eq:a} and \eqref{eq:Qo}, respectively. 
Their formal expressions are provided by
\begin{alignat*}{3}
	&\nabla_{\mathbf{x}}l^{fo1} &&= \begin{bmatrix}
		\nabla^{\top}_{\mathbf{p}}l^{fo1} & \nabla^{\top}_{\dot{\mathbf{p}}}l^{fo1}
	\end{bmatrix}^{\top} &&= \begin{bmatrix}
		\nabla^{\top}_{\mathbf{p}}l^{fo1} & \mathbf{0}^{\top}_{N}
	\end{bmatrix}^{\top}\\
	&\mathcal{H}_{\mathbf{x}\mathbf{x}} l^{fo1} &&= \begin{bmatrix}
		\mathcal{H}_{\mathbf{p}\mathbf{p}} l^{fo1} & \mathcal{H}_{\mathbf{p}\dot{\mathbf{p}}} l^{fo1} \\
		\mathcal{H}_{\dot{\mathbf{p}}\mathbf{p}} l^{fo1} & \mathcal{H}_{\dot{\mathbf{p}}\dot{\mathbf{p}}} l^{fo1}
	\end{bmatrix} &&= \begin{bmatrix}
		\mathcal{H}_{\mathbf{p}\mathbf{p}} l^{fo1} &  \Zz_{N}  \\
		\Zz_{N}								  &  \Zz_{N}
	\end{bmatrix}
\end{alignat*}
where, by assigning $\boldsymbol{\Pi}_{ij}(t)=\mathbf{e}_{ij}(t) \mathbf{e}_{ij}(t)^{\top} \in \mathbb{R}^{M\times M}$, it holds that, for all $i=1,\dots,n$,
\begin{equation*}\label{eq:gradfo}
	\nabla_{\mathbf{p}_{i}}l^{fo1} = k_{F} \sum\limits_{j \in \mathcal{N}_{i}} \sigma_{d_{ij}}^{\prime}   \mathbf{e}_{ij}
\end{equation*}
and, for all $(i,j) \in \mathcal{E}$,
\begin{equation}\label{eq:hessfoout}
	\mathcal{H}_{\mathbf{p}_{i}\mathbf{p}_{j}} l^{fo1} = -k_{F}\left(2\sigma_{d_{ij}} ^{\prime\prime}   \boldsymbol{\Pi}_{ij} + \sigma_{d_{ij}}^{ \prime}   \mathbf{I}_{M} \right);
\end{equation}
otherwise, for $(i,j) \notin \mathcal{E}$ one has $\mathcal{H}_{\mathbf{p}_{i}\mathbf{p}_{j}} l^{fo1} = \Zz_{M}$.
Notably, condition $\sigma_{d_{ij}} \in \mathscr{C}^{2}$, for all $(i,j)\in\mathcal{E}$, is required to use PRONTO, in order for \eqref{eq:hessfoout} not to present discontinuities. An expression for the diagonal blocks in the Hessian $\mathcal{H}_{\mathbf{p}\mathbf{p}} l^{fo1}$ is then provided by
\begin{equation}\label{eq:hessfoin}
	\mathcal{H}_{\mathbf{p}_{i}\mathbf{p}_{i}} l^{fo1} = - \sum\limits_{\forall j\neq i} \mathcal{H}_{\mathbf{p}_{i}\mathbf{p}_{j}} l^{fo1}.
\end{equation}
Similarly, the gradient and Hessian matrix of term $l^{fo2}$ in \eqref{eq:a}-\eqref{eq:Qo} are yielded by
\begin{alignat*}{3}
	&\nabla_{\mathbf{x}}l^{fo2} &&= \begin{bmatrix}
		\nabla^{\top}_{\mathbf{p}}l^{fo2} & \nabla^{\top}_{\dot{\mathbf{p}}}l^{fo2}
	\end{bmatrix}^{\top} &&= \begin{bmatrix}
		\mathbf{0}^{\top}_{N} & \nabla^{\top}_{\dot{\mathbf{p}}}l^{fo2}
	\end{bmatrix}^{\top}\\
	&\mathcal{H}_{\mathbf{x}\mathbf{x}} l^{fo2} &&= \begin{bmatrix}
		\mathcal{H}_{\mathbf{p}\mathbf{p}} l^{fo2} & \mathcal{H}_{\mathbf{p}\dot{\mathbf{p}}} l^{fo2} \\
		\mathcal{H}_{\dot{\mathbf{p}}\mathbf{p}} l^{fo2} & \mathcal{H}_{\dot{\mathbf{p}}\dot{\mathbf{p}}} l^{fo2}
	\end{bmatrix} &&= \begin{bmatrix}
		\Zz_{N}                                &  \Zz_{N}  \\
		\Zz_{N}								  &  \mathcal{H}_{\dot{\mathbf{p}}\dot{\mathbf{p}}} l^{fo2}
	\end{bmatrix}
\end{alignat*}
where 
\begin{equation*}\label{eq:gradal}
	\nabla_{\dot{\mathbf{p}}_{i}}l^{fo2}(t) = k_{A} \sum\limits_{ j \in \mathcal{N}_{i}} \Thth_{ij} \dot{\mathbf{e}}_{ij} 
\end{equation*}
and for all $(i,j) \in \mathcal{E}$,
\begin{equation*}\label{eq:hessal}
	\mathcal{H}_{\dot{\mathbf{p}}_{i}\dot{\mathbf{p}}_{j}} l^{fo2} = 
	-k_{A} \Thth_{ij};
\end{equation*}
otherwise, for $(i,j) \notin \mathcal{E}$ one has $\mathcal{H}_{\dot{\mathbf{p}}_{i}\dot{\mathbf{p}}_{j}} l^{fo2} = \Zz_{M}$. Also, \begin{equation*}\label{eq:hessfal}
	\mathcal{H}_{\dot{\mathbf{p}}_{i}\dot{\mathbf{p}}_{i}} l^{fo2} = - \sum\limits_{\forall j \neq i}  \mathcal{H}_{\dot{\mathbf{p}}_{i}\dot{\mathbf{p}}_{j}} l^{fo2}.
\end{equation*}

As a matter of fact, 
PRONTO involves the resolution of an LQ problem to compute optimal search directions. Hence, $\mathbf{Q}_{o}(t)$ should be taken positive semidefinite for all $t \geq 0$.
In fact, it can be observed that expression \eqref{eq:Qo} does not always guarantee this condition, given the general undetermined definiteness of the Hessian related to the formation $\mathcal{H}_{\mathbf{p}\dot{\mathbf{p}}} l^{fo1}$. To this aim, a safe version of $\mathbf{Q}_{o}$, say $\mathbf{Q}^{safe}_{o}$, is implemented by exploiting a heuristic based on the Gershgorin circle theorem \cite{Bell1965}. This is carried out in practice by computing for $i\neq j$ the off-diagonal blocks
\begin{equation*}\label{eq:heuristicsPronto}
	\mathcal{H}^{safe}_{\mathbf{p}_{i}\mathbf{p}_{j}} l^{fo1} = -k_{F} \left(2\sigma_{d_{ij}} ^{\prime\prime}   \boldsymbol{\Pi}_{ij} + 
	\chi_{0}(\sigma_{d_{ij}}^{ \prime}) \sigma_{d_{ij}}^{ \prime}   \mathbf{I}_{M} \right), 
\end{equation*}
where $\chi_{0}:\mathbb{R}\rightarrow \{0,1\}$ denotes the unitary characteristic function such that it returns the value $0$ if its argument is negative; the value $1$, otherwise. The diagonal blocks for such a safe version of $\mathcal{H}_{\mathbf{p}\mathbf{p}} l^{fo1}$ are then obtained exploiting \eqref{eq:hessfoin}. 

\subsection{Preliminary discussion for the design of a decentralized controller}
In the previous section, the main results obtained in \cite{FabrisCenedeseHauser2019} have been re-elaborated to solve the generalized version of the OIFT problem \eqref{eq:minimizationproblemOIFT}-\eqref{eq:align_inst_cost}. However, it has to be noticed that the PRONTO inverse dynamics cannot be applied in an online fashion and its computation is centralized.
Henceforward, we build up the analytical reasoning towards a distributed control law $\widehat{\uu}(t) = \vci{\widehat{\uu}_{i}(t)}{n} $ that solves the OIFT problem in the finite-time interval $[0,T]$. 
To this purpose, we list  the three challenging design criteria that will be observed to derive a suitable decentralized controller: 
\begin{enumerate}
	\item[$(i)$] $\widehat{\mathbf{u}}$ must comply with optimality criteria in order to solve problem \eqref{eq:minimizationproblemOIFT}, possibly, resorting to heuristics;
	\item[$(ii)$] $\widehat{\mathbf{u}}$ must be distributed, namely $\widehat{\mathbf{u}}_{i}$ is computed only with the information of nodes $j \in \overline{\mathcal{N}}_{i} = \mathcal{N}_{i} \cup \{i\}$;
	\item[$(iii)$] $\widehat{\mathbf{u}}$ must be an online feedback control law, i.e. it can be implemented in a real-time device (assuming to have enough hardware capabilities) by leveraging the information of the current global state of the system.
\end{enumerate}

The derivation of $\widehat{\uu}$ starts by tackling point $(i)$, specifically. To this aim, we first provide the next preliminary definition.
\begin{Def}[Hamiltonian associated to \eqref{eq:dyn_sys}-\eqref{eq:minimizationproblemOIFT}]
	Let $\boldsymbol{\lambda}(t) = \vci{\boldsymbol{\lambda}_{i}(t)}{2n} \in \mathbb{R}^{2N}$ be the time-varying co-state vector, where $\boldsymbol{\lambda}_{i} \in \mathbb{R}^{M}$. The real scalar function 
	\begin{equation}
		H = \boldsymbol{\lambda}^{\top} \mathbf{f} + l = \boldsymbol{\lambda}^{\top}(\mathbf{A}\mathbf{x}+\mathbf{B}\mathbf{u}) + l^{st}(\mathbf{x}) + l^{in}(\mathbf{u})
	\end{equation} 
	is said to be the \textit{Hamiltonian} associated to problem \eqref{eq:minimizationproblemOIFT} constrained to dynamics \eqref{eq:dyn_sys}.
\end{Def}

Variables $\boldsymbol{\lambda}_{i}$, for $i=1,\ldots,2n$, are referred to as the \textit{Lagrangian multipliers}.
Now, to make $\widehat{\uu}$ comply with optimality criteria, a weak version of the Pontryagin Minimum Principle (PMP) \cite{Pontryagin2018} applied to the specific framework under analysis is provided in the next theorem. 

\begin{theorem}[PMP applied to \eqref{eq:dyn_sys}-\eqref{eq:minimizationproblemOIFT}] \label{lem:PMP}
	Let $\mathcal{U}\subseteq\mathbb{R}^{N}$ be the set of all admissible inputs and $\mathbf{u}^{\star}(t) \in \mathcal{U}$ be a particular input for dynamics \eqref{eq:dyn_sys}. The necessary conditions for $\mathbf{u}^{\star}$ to solve problem \eqref{eq:minimizationproblemOIFT} are yielded by
	\begin{equation}\label{eq:PMPstates}
		\begin{cases}
			\dot{\mathbf{x}} = \nabla_{\boldsymbol{\lambda}} H \vert_{\mathbf{u}=\mathbf{u}^{\star}} \\
			\mathbf{x}(0) = \mathbf{x}_{0}	
		\end{cases} \Leftrightarrow ~~
		\begin{cases}
			\dot{\mathbf{x}} = \mathbf{A}\mathbf{x}+\mathbf{B}\mathbf{u}^{\star} \\
			\mathbf{x}(0) = \mathbf{x}_{0}	
		\end{cases},
	\end{equation}
	\begin{equation}\label{eq:PMPcostates}
		\begin{cases}
			-\dot{\boldsymbol{\lambda}} = \nabla_{\mathbf{x}} H \vert_{\mathbf{u}=\mathbf{u}^{\star}} \\
			\boldsymbol{\lambda}(T) = \nabla_{\mathbf{x}} m(\mathbf{x}(T))
		\end{cases} \!\!\!\Leftrightarrow ~
		\begin{cases}
			-\dot{\boldsymbol{\lambda}} = \mathbf{A}^{\top}\boldsymbol{\lambda} + \nabla_{\mathbf{x}} l^{st}(\mathbf{x}) \\
			\boldsymbol{\lambda}(T) = \nabla_{\mathbf{x}} m(\mathbf{x}(T))
		\end{cases}
	\end{equation}
	and
	\begin{equation}\label{eq:PMPinput}
		\nabla_{\mathbf{u}} H \vert_{\mathbf{u}=\mathbf{u}^{\star}} = 0 ~~\Leftrightarrow ~~ \mathbf{u}^{\star} = -\mathbf{R}^{-1} \mathbf{B}^{\top} \boldsymbol{\lambda}.
	\end{equation}
	Denoting with $(\xx^{\star}(\cdot),\lamlam^{\star}(\cdot))$ the evolution of $(\xx,\lamlam)$ subject to input $\uu^{\star}$, conditions \eqref{eq:PMPstates}-\eqref{eq:PMPinput} imply that $\uu^{\star}$ is a candidate to satisfy
	\begin{equation}
		H(\xx^{\star},\lamlam^{\star},\uu^{\star}) \leq H(\xx^{\star},\lamlam^{\star},\uu), \quad \forall \uu \in \mathcal{U}.
	\end{equation}
	Moreover, assuming that an input $\mathbf{u}^{\star}$ satisfies \eqref{eq:PMPstates}-\eqref{eq:PMPinput}, the following condition is sufficient for $\mathbf{u}^{\star}(t)$ to be an instantaneous minimizer at time $t$:
	\begin{equation}\label{eq:PMPsuff}
		\begin{bmatrix}
			\mathcal{H}_{\mathbf{x}\mathbf{x}}  H(t) & \mathcal{H}_{\mathbf{x}\mathbf{u}}  H(t) \\
			\mathcal{H}_{\mathbf{u}\mathbf{x}}  H(t) & \mathcal{H}_{\mathbf{u}\mathbf{u}}  H(t) 
		\end{bmatrix} \succeq 0 ~\Leftrightarrow\! ~
		\begin{cases}
			\mathcal{H}_{\mathbf{x}\mathbf{x}}  l^{st}(t) \succeq 0 \\
			\mathbf{R} \succeq 0
		\end{cases}.
	\end{equation}
	If \eqref{eq:PMPsuff} holds for all $t\in [0,T]$ then $\mathbf{u}^{\star}$ solves problem \eqref{eq:minimizationproblemOIFT}.
\end{theorem}
\begin{proof}
	The statement is a direct consequence of the PMP applied to \eqref{eq:dyn_sys}-\eqref{eq:cost}, as shown by each double implication in \eqref{eq:PMPstates}-\eqref{eq:PMPinput} and \eqref{eq:PMPsuff}. In particular, the sufficient condition \eqref{eq:PMPsuff} is equivalent to require convexity for $l$ w.r.t. variables $\mathbf{x}$ and $\boldsymbol{\lambda}$, leveraging the fact that \eqref{eq:dyn_sys} is linear (see also \cite{Locatelli2001}).
\end{proof}

The PMP of Thm. \ref{lem:PMP} will be exploited to a certain degree in Sec. \ref{sec:distr_law} with the aim of characterizing the structure of the proposed online distributed control law.

We now address point $(ii)$. To this purpose, the derivation of a local estimator $\widehat{\mathbf{x}}_{c}^{i}(t)$, for each agent $i=1,\dots,n$, that reconstructs the information about the centroid $\mathbf{x}_{c}(t)$ will be crucial, since one term of controller $\uu^{\star}$ is devoted to satisfy the minimization of the tracking part of $h$ influenced by $l^{tr}$. Such estimator $\widehat{\mathbf{x}}_{c}^{i}$ is of second-order type, as it mimics dynamics \eqref{eq:dyn_sys}. 
The next proposition thus focuses on bounding the estimation error for the
first-order estimator devised by the authors of \cite{AntonelliArricchielloCaccavale2013}. This result will be subsequently employed to derive a second-order estimator having the purpose to recover $\widehat{\mathbf{x}}_{c}^{i}$ in a decentralized fashion. 

\begin{prop}[Local first-order centroid estimator] \label{prop:loccentrest}
	Given an undirected and connected graph $\mathcal{G}=(\mathcal{V},\mathcal{E})$, assume that the dynamics of the whole system of $n$ agents associated to $\mathcal{G}$ is described by the single integrator
	\begin{equation}
		\dot{\mathbf{y}}(t) = \mathbf{w}(t)
	\end{equation}
	where $\mathbf{y}= \vci{\yy_{i}}{n}\in \mathbb{R}^{N}$, $\mathbf{w}=\vci{\ww_{i}}{n} \in \mathbb{R}^{N}$, $N=nM$, and with $\yy_{i},\ww_{i}\in \mathbb{R}^{M}$ denoting the state-input couple for the $i$-th agent.
 	In addition, for $i=1,\dots,n$, let $\mathbf{N}_{i} \in \mathbb{R}^{N\times N}$ be the projection matrix such that for a vector $\psps \in \mathbb{R}^{N}$, $\psps = \vci{\psps_{i}}{n}$, with $\psps_{i} \in \mathbb{R}^{M}$, 
 	it holds that $\mathbf{N}_{i} \psps = \begin{bmatrix}
		\mathbf{0}_{M}^{\top} &\cdots & \mathbf{0}_{M}^{\top} & \psps_{i}^{\top} & \mathbf{0}_{M}^{\top} & \cdots & \mathbf{0}_{M}^{\top}
	\end{bmatrix}^{\top}$.
	Let us consider the $i$-th first-order state-space estimator
	\begin{equation}\label{eq:ssestim}
		\begin{cases}
			\dot{\widehat{\mathbf{y}}}^{i} = -k_{y} \sum\limits_{j \in \mathcal{N}_{i}} (\widehat{\mathbf{y}}^{i}-\widehat{\mathbf{y}}^{j}) -k_{y} \mathbf{N}_{i}(\widehat{\mathbf{y}}^{i}-\mathbf{y}) + \widehat{\mathbf{w}}^{i} \\
			\widehat{\mathbf{y}}_{c}^{i} = n^{-1} \sum_{j=1}^{n}\widehat{\mathbf{y}}^{i}_{j}
   , \quad \widehat{\mathbf{y}}^{i}(0) = \widehat{\mathbf{y}}^{i}_{0}
		\end{cases},
	\end{equation}
	such that $k_{y}$ is a positive gain, $\widehat{\mathbf{y}}^{i}_{0} \in \mathbb{R}^{N}$ is an arbitrary initial condition, and $\widehat{\yy}^{i} = \vcj{\widehat{\yy}_{j}^{i}}{n} \in \mathbb{R}^{N}$, $\widehat{\ww}^{i} = \vcj{\widehat{\ww}_{j}^{i}}{n} \in \mathbb{R}^{N}$ for $i = 1,\ldots,n$, with $\widehat{\yy}_{j}^{i}, \widehat{\ww}_{j}^{i} \in \mathbb{R}^{M}$. Assume that the $j$-th component of the $i$-th input is selected as 
	\begin{equation}\label{eq:inputestassignment}
		\widehat{\mathbf{w}}^{i}_{j}(t) = \begin{cases}
			\ww_{j}(t), ~\text{if } j\in \overline{\mathcal{N}}_{i}; \\
			\widetilde{\ww}_{j}^{i}(t),  ~\text{otherwise};
		\end{cases}
	\end{equation}
	where $\widetilde{\ww}_{j}^{i} \in \mathbb{R}^{M}$ has to be designed.
	Furthermore, let us define the $i$-th centroid estimation error $\ee_{y_{c}}^{i} = \mathbf{y}_{c}- \widehat{\mathbf{y}}_{c}^{i}$ between the actual centroid $\mathbf{y}_{c} = n^{-1} \sum_{i=1}^{n} \yy_{i}$ and its $i$-th estimate $\widehat{\mathbf{y}}_{c}^{i}$. 
	If for all $i = 1,\ldots,n$
	\begin{equation}\label{eq:hypconvinpest}
		\underset{t\rightarrow +\infty}{\lim} \ww_{j}(t)-\widetilde{\ww}_{j}^{i}(t) = \mathbf{g}_{ij}, ~ \forall j \in \mathcal{V} \setminus \overline{\mathcal{N}}_{i},
	\end{equation}
	with $\mathbf{g}_{ij} \in \mathbb{R}^{M}$ finite constant, then, denoting $\ee_{y_{c}} = \vci{\ee^{i}_{y_{c}}}{n}$, one has
	\begin{equation}\label{eq:centrerrest}
		\left\|\ee_{y_{c}}(t)\right\|_{2}^{2} \leq \varpi_{y_{c}}^{2}(t) := \dfrac{\mathfrak{c}(\mathcal{G})^{2}}{n^{2}k_{y}^{2}} \sum\limits_{i=1}^{n} \sum\limits_{j \in \mathcal{V} \setminus \overline{\mathcal{N}}_{i} } \left\|\ww_{j}(t)-\widetilde{\ww}_{j}^{i}(t) \right\|^{2}_{2}
	\end{equation}
	as $t\rightarrow +\infty$, where $\mathfrak{c}(\mathcal{G}) > 0$ is a topological constant depending only on $\mathcal{G}$.
\end{prop}
\begin{proof}
	This result is proven showing that the $i$-th estimate $\widehat{\mathbf{y}}^{i}$ converges exponentially fast towards $\mathbf{y}$ as $t \rightarrow +\infty$. Let us define $\ee_{y}^{i} = \yy-\widehat{\yy}^{i}$ and $\ee_{w}^{i} = \ww-\widehat{\ww}^{i}$ as the state and input errors for system \eqref{eq:ssestim}, respectively. The dynamics of each $\ee_{y}^{i}$ is then described by
	\begin{equation}
		\dot{\ee}_{y}^{i} = -k_{y}  \sum\limits_{j \in \mathcal{N}_{i}} (\ee_{y}^{i}-\ee_{y}^{j}) -k_{y} \Nn_{i}\ee_{y}^{i} + \ee_{w}^{i}.
	\end{equation}
	Now, defining the global state error $\ee_{y} = \vci{\ee_{y}^{i}}{n}$, the global input error $\ee_{w} = \vci{\ee_{w}^{i}}{n}$ and resorting to the Laplacian $\Ll$ of graph $\mathcal{G}$ one has 
	\begin{equation}\label{eq:errdynest}
		\dot{\ee}_{y} = -k_{y} \left[(\Ll \otimes \Ii_{N}
  ) + \Nn \right] \ee_{y} + \ee_{w}
	\end{equation}
	where $\Nn = \Diag(\Nn_{i})$ and $\otimes$ denotes the Kronecker product. Indicating the update matrix of \eqref{eq:errdynest} with $\Mm = -k_{y} \left[(\Ll \otimes \Ii_{N}
 ) + \Nn \right] \in \mathbb{R}^{{nN \times nN}}
 $, it is possible to observe that $\rk(\Mm)$ is full. This is due by the fact that, in $\Mm$, the diagonal projection matrix $\Nn$, with $\rk(\Nn) = N$, adds to the extended Laplacian $(\Ll \otimes \Ii_{N}
 )$ (for more details, see also \cite{AntonelliArricchielloCaccavale2013}). 
 The matrix $\Mm$ is indeed positive semidefinite: it has 
 exactly $N$ zero eigenvalues by the Kronecker product properties and thus $\rk(\Ll \otimes \Ii_{N}
 ) = (n-1)N
 $. As a consequence, the diagonal elements of $\Nn$ contribute to form $N$
 self loops in the graph associated to $(\Ll \otimes \Ii_{N}
 )$; hence, $\rk(\Mm) = nN 
 $ is achieved. Moreover, since matrix $\Nn$ is also positive semidefinite and $\rk(\Mm) = nN
 $, matrix $\Mm$ is negative definite. Therefore, by also leveraging hypothesis \eqref{eq:hypconvinpest}, the linear dynamics in \eqref{eq:errdynest} stabilizes around
	\begin{equation}
		\ee_{y}^{\star}(t) = k_{y}^{-1} \left[(\Ll \otimes \Ii_{N}
  ) + \Nn \right]^{-1} \ee_{w}(t) = -\Mm^{-1} \ee_{w}(t)
	\end{equation} 
	for large values of $t$. 
	The latter convergence result can be proven by choosing the Lyapunov function $V(\ee_{y}) = (\ee_{y}-\ee_{y}^{\star})^{\top}(\ee_{y}-\ee_{y}^{\star})/2$ and showing that $\dot{V}(\ee_{y}) <0$ as $t\rightarrow +\infty$. In fact, denoting with $\mathbf{g} \in \mathbb{R}^{nN}$ the vector whose components are the constants $\mathbf{g}_{ij}$ so that $\mathbf{g}$ is the value taken by $\ee_{w}(t)$ as $t\rightarrow +\infty$, it follows that $\dot{V}(\ee_{y}) = (\Mm\ee_{y}+\mathbf{g})^{\top}\Mm^{-1}(\Mm\ee_{y}+\mathbf{g}) < 0$ for all $\ee_{y} \neq -\Mm^{-1}\mathbf{g}$.
	In conclusion, the following inequality can be derived:
	\begin{equation}\label{eq:ineqerrbase}
		\left\|\ee_{y}\right\|_{2} = k_{y}^{-1} \left\| \left[(\Ll \otimes \Ii_{N}
  ) + \Nn \right]^{-1} \ee_{w}\right\|_{2} \leq \dfrac{\mathfrak{c}(\mathcal{G})}{k_{y}} \left\|\ee_{w}\right\|_{2}
	\end{equation}
	where constant $\mathfrak{c}(\mathcal{G}) > 0$ is the spectral radius of $\left[(\Ll \otimes \Ii_{N}
 ) + \Nn \right]^{-1} \succ 0$. From
	\begin{align}\label{eq:errycij}
		\left\|\ee_{y_{c}}^{i}\right\|_{2}^{2} &= \left\|\dfrac{1}{n} \sum\limits_{j=1}^{n} \ee_{y_{j}}^{i} \right\|_{2}^{2} \leq \dfrac{1}{n^{2}}\sum\limits_{j=1}^{n}\left\| \ee_{y_{j}}^{i}  \right\|_{2}^{2}
	\end{align}
	and \eqref{eq:ineqerrbase} it follows that
	\begin{align}
		\left\|\ee_{y_{c}}\right\|_{2}^{2} &= \sum\limits_{i=1}^{n} \left\|\ee_{y_{c}}^{i}\right\|_{2}^{2} \leq  \dfrac{1}{n^{2}}\sum\limits_{i=1}^{n}\sum\limits_{j=1}^{n} \left\| \ee_{y_{j}}^{i}  \right\|_{2}^{2}  \\
		&= \dfrac{1}{n^{2}} \left\|\ee_{y}\right\|_{2}^{2} \leq  \dfrac{\mathfrak{c}(\mathcal{G})^{2}}{n^{2}k_{y}^{2}} \left\|\ee_{w}\right\|_{2}^{2} =\varpi_{y_{c}}^{2}(t)
	\end{align}
	as $t\rightarrow +\infty$ and thus \eqref{eq:centrerrest} is finally proven. 
\end{proof}

\begin{rmk}\label{rm:desiredinputest}
	In Prop. \ref{prop:loccentrest}, dynamics \eqref{eq:ssestim} represents the basis to construct an estimator that computes estimations of both position and velocity centroids $(\pp_{c},\dot{\pp}_{c})$ at each node $i = 1,\ldots,n$. In particular, terms $\widetilde{\ww}_{j}^{i}$ in \eqref{eq:inputestassignment} are assigned according to desired values for the input components $\ww_{j}$ that are not available to agent $i$. It is to note that, if $\ww_{j}$ meets the choice of $\widetilde{\ww}_{j}^{i}$ as expected when $t \rightarrow +\infty$, for any agent $i$, condition \eqref{eq:centrerrest} guarantees that the estimation of the centroid $\yy_{c}$ is accurate, namely the related estimation error $\ee_{y_{c}}$ vanishes as $t \rightarrow +\infty$ (we will see that this condition is satisfied in the framework under analysis whenever the control input $\uu$ tends to zero). In addition, it is worth observing that: (1) high values for $k_{y}$ reduce the upper bound in \eqref{eq:centrerrest};
	(2) the more $\mathcal{G}$ is close to be regular bipartite, the more $\mathfrak{c}(\mathcal{G})$ increases and, for a fixed $n$, $\mathfrak{c}(\mathcal{G})$ decreases as $|\mathcal{E}|$ decreases (see \cite{Shi2007}).
\end{rmk}

We now propose a local second-order estimator
\begin{equation}
	\widehat{\mathbf{x}}_{c}^{i}(t) = \vc{\widehat{\mathbf{p}}_{c}^{i}(t),\widehat{\dot{\mathbf{p}}}_{c}^{i}(t)} \in \mathbb{R}^{2N}, \quad i = 1,\ldots,n
\end{equation}
in the following corollary. 

\begin{corollary}[Local second-order centroid estimator]\label{lem:secordest}
	Let $k_{py}$ and $k_{dy}$ be two positive gains. Each node $i = 1,\ldots,n$ can compute the $i$-th estimates $(\widehat{\pp}_{c}^{i},\widehat{\dot{\pp}}_{c}^{i}) = \left(n^{-1}\sum_{j=1}^{n}\widehat{\pp}_{j}^{i}, n^{-1}\sum_{j=1}^{n}\widehat{\dot{\pp}}_{j}^{i} \right)$ for the position and velocities centroids $(\pp_{c},\dot{\pp}_{c})$ in the OIFT problem \eqref{eq:minimizationproblemOIFT} constrained to dynamics \eqref{eq:dyn_sys} exploiting the $i$-th second-order estimator
	\begin{equation}\label{eq:estimator2}
		\begin{cases}
			\dfrac{\mathrm{d}}{\mathrm{d}t}\widehat{\pp}^{i} = -k_{py}  \sum\limits_{j\in \mathcal{N}_{i}} (\widehat{\pp}^{i}-\widehat{\pp}^{j}) -k_{py} \mathbf{N}_{i} (\widehat{\pp}^{i}-\pp ) + \widehat{\dot{\pp}}^{i} \\ 
			\dfrac{\mathrm{d}}{\mathrm{d}t}\widehat{\dot{\pp}}^{i} = -k_{dy}  \sum\limits_{j\in \mathcal{N}_{i}} (\widehat{\dot{\pp}}^{i}-\widehat{\dot{\pp}}^{j}) -k_{dy} \mathbf{N}_{i} (\widehat{\dot{\pp}}^{i}-\dot{\pp} ) + \widehat{\uu}^{i}
		\end{cases}
	\end{equation}
	such that $\widehat{\pp}^{i} = \vcj{\widehat{\pp}^{i}_{j}}{n}$, $\widehat{\dot{\pp}}^{i} = \vcj{\widehat{\dot{\pp}}^{i}_{j}}{n}$, $\widehat{\uu}^{i} = \vcj{\widehat{\uu}^{i}_{j}}{n}$, with
	\begin{equation}\label{eq:estinupt2ord}
			\widehat{\uu}^{i}_{j} = \begin{cases}
				\uu_{j}, &\text{if } j\in \overline{\mathcal{N}}_{i}; \\
				\mathbf{0}_{M}, &\text{otherwise};
			\end{cases}
		\end{equation}
	and initialized as\footnote{In the second row of \eqref{eq:initializationofestimator}, if the information on the degree $|\mathcal{N}_{k}|$ is not available one can simply average $\vc{\pp_{k}(0),\dot{\pp}_{k}(0)}$ over $|\overline{\mathcal{N}}_{i}|$.}
	\begin{equation}\label{eq:initializationofestimator}
		\begin{bmatrix}
			\widehat{\pp}^{i}_{j}(0) \\ \widehat{\dot{\pp}}^{i}_{j}(0)
		\end{bmatrix} = \begin{cases}
			\vc{\pp_{j}(0),\dot{\pp}_{j}(0)}, &\text{if } j\in \overline{\mathcal{N}}_{i}; \\  \dfrac{\sum\limits_{k\in \overline{\mathcal{N}}_{i}}(1+|\mathcal{N}_{k}|)\vc{\pp_{k}(0),\dot{\pp}_{k}(0)}}{\sum_{k\in \overline{\mathcal{N}}_{i}} (1+|\mathcal{N}_{k}|)}, &\text{otherwise}.
		\end{cases}
	\end{equation}
	
	Furthermore, if $\mathbf{u}$ is a stabilizing input for system \eqref{eq:dyn_sys} and is such that $\mathbf{u} \rightarrow \mathbf{0}_{N}$ as $t$ grows then, for large values of $t$, 
	the $i$-th position estimation error $\ee_{p}^{i} = \pp - \widehat{\pp}^{i} $ and the $i$-th velocity estimation error $\ee_{\dot{p}}^{i} =  \dot{\pp} - \widehat{\dot{\pp}}^{i} $ vanish. In particular, setting $\ee_{p_{c}}^{i} = n^{-1} \sum_{j=1}^{n} \ee_{p_{j}}^{i}$, $\ee_{\dot{p_{c}}}^{i} = n^{-1} \sum_{j=1}^{n} \ee_{\dot{p}_{j}}^{i}$, $\ee_{p_{c}} = \vci{\ee_{p_{c}}^{i}}{n}$, $\ee_{\dot{p_{c}}} = \vci{\ee_{\dot{p_{c}}}^{i}}{n}$ one has
	\begin{equation}\label{eq:ineqerrbase2}
		\left\|	\ee_{p_{c}}(t) \right\|_{2}^{2} + \left\|	\ee_{\dot{p}_{c}}(t) \right\|_{2}^{2}  \leq  \dfrac{\mathfrak{c}(\mathcal{G})^2}{n^{2} \min(k_{py},k_{dy})^{2}} \sum\limits_{i=1}^{n} \sum\limits_{j\in \mathcal{V}\setminus \overline{\mathcal{N}}_{i}} \left\|\uu_{j}(t) \right\|^{2}_{2}
	\end{equation}
	as $t$ gets larger if each $\uu_{j}^{i}(t)$ tends to some finite value.
\end{corollary}
\begin{proof}
	The estimator developed in this lemma is the natural second-order extension of the results illustrated in Prop. \ref{prop:loccentrest}. One can demonstrate the convergence properties for the estimation errors by analyzing the stability of the system
	\begin{equation}\label{eq:errordynamicssoe}
		\begin{bmatrix}
			\dot{\ee}_{p} \\ \dot{\ee}_{\dot{p}} 
		\end{bmatrix} =
		\begin{bmatrix}
			\Mm_{p} & \Ii_{Nn} \\ \mathbf{O}_{Nn} & \Mm_{\dot{p}} 
		\end{bmatrix}
		\begin{bmatrix}
			\ee_{p} \\ \ee_{\dot{p}} 
		\end{bmatrix} +
		\begin{bmatrix}
			\mathbf{0}_{Nn} \\ \ee_{u} 
		\end{bmatrix}
	\end{equation}
	where $\dot{\ee}_{p} = \vci{\dot{\ee}_{p}^{i}}{n}$, $\dot{\ee}_{\dot{p}} = \vci{\dot{\ee}_{\dot{p}}^{i}}{n}$, $\ee_{u} =  \vci{\uu-\widehat{\uu}^{i}}{n}$, $\Mm_{p} = -k_{py} ((\Ll \otimes \Ii_{N})+\Nn)$, $\Mm_{\dot{p}} = -k_{dy} ((\Ll \otimes \Ii_{N})+\Nn)$ and with $\Nn$ defined as in Prop. \ref{prop:loccentrest}. Given the structure of the update matrix in \eqref{eq:errordynamicssoe}, hereafter denoted with $\overline{\Mm}$, it can be noted that its eigenvalues exactly coincide with those of $\Mm_{p}$ and $\Mm_{\dot{p}}$. Using the same reasoning in the proof of Prop. \ref{prop:loccentrest}, the system in \eqref{eq:errordynamicssoe} stabilizes around $\ee^{\star}(t) = -\overline{\Mm}^{-1} \begin{bmatrix}
		\mathbf{0}_{Nn}^{\top} & \ee_{u}(t)^{\top}
	\end{bmatrix}^{\top} $ as $t$ grows.
	As also pointed out in Rem. \ref{rm:desiredinputest}, the exogenous inputs for the estimator chosen in \eqref{eq:estinupt2ord} (i.e., the entries $\uu_{j}$ if $j\in \overline{\mathcal{N}}_{i}$; $\mathbf{0}_{M}$, otherwise) are selected in such a way that both $\ee_{p}^{i}$  and $\ee_{\dot{p}}^{i}$ vanish for large values of $t$.\\
	Finally, relation \eqref{eq:ineqerrbase2} can be derived similarly to \eqref{eq:ineqerrbase} by leveraging the 
	dynamics \eqref{eq:errordynamicssoe} and adopting \eqref{eq:estinupt2ord}.
	
\end{proof}

\subsection{Online distributed feedback control law}\label{sec:distr_law}
In light of the above reasoning, we are now ready to discuss the distributed feedback control law complying with the design criteria $(i)$-$(iii)$.
To this aim, we propose the distributed feedback control law $\widehat{\uu} = \vci{\widehat{\uu}_{i}}{n} $, $\widehat{\uu}_{i} \in \mathbb{R}^{M}$, that only relies on each agent's local reference frame and is defined as
\begin{align}\label{eq:distrOIFTsol}
	\widehat{\uu}_{i} =& -\mathbf{R}_{i}^{-1}(k_{p}^{tr1} \nabla_{\pp_{i}} l^{tr1}+k_{d}^{tr1} \dot{\nabla}_{\pp_{i}} l^{tr1} +  k_{p}^{fo1} \nabla_{\pp_{i}} l^{fo1}+ \nonumber\\
	&k_{d}^{fo1} \dot{\nabla}_{\pp_{i}}^{safe} l^{fo1} + k_{p}^{tr2} \nabla_{\dot{\pp}_{i}} l^{tr2}  + k_{p}^{fo2} \nabla_{\dot{\pp}_{i}} l^{fo2}) ,
\end{align}
where terms $ \nabla_{\pp_{i}} l^{tr1}, \dot{\nabla}_{\pp_{i}} l^{tr1}, \nabla_{\dot{\pp}_{i}} l^{tr2} $ in \eqref{eq:distrOIFTsol} are computed through the second-order local estimator in Cor. \ref{lem:secordest}. 
Constants $k_{p}^{tr1},k_{p}^{fo1},k_{d}^{tr1},k_{d}^{fo1},k_{p}^{tr2},$ $k_{p}^{fo2}$ are positive control gains that can be adjusted by the designer, e.g. by matching the centralized solution during simulations and considering that law in \eqref{eq:distrOIFTsol} has a structure similar to a PD controller\footnote{Such control gains cannot be generally tuned a-priori, as their values depend on the specific network interconnections and the chosen parameter weights.}. 

The expression of the control law in \eqref{eq:distrOIFTsol} can be justified as follows.
The design criterion $(i)$ is satisfied, since the structure of $\widehat{\uu} = -\Rr^{-1}\Bb^{\top} \widehat{\lamlam}$ mimics the one provided by the optimal control $\uu^{\star} = -\Rr^{-1}\Bb^{\top} \lamlam$ in Thm. \ref{lem:PMP}. Indeed, the solution of the co-state equation in \eqref{eq:PMPcostates} can be reproduced by means of
\begin{equation}\label{eq:lamdynest}
	\forall i=1,\ldots,n:~~\begin{cases}
		\widehat{\lamlam}_{i} =  k_{p}^{tr1} \nabla_{\pp_{i}} l^{tr1}+k_{d}^{tr1} \dot{\nabla}_{\pp_{i}} l^{tr1} + \\ ~~~~~~~\! k_{p}^{fo1} \nabla_{\pp_{i}} l^{fo1} + k_{d}^{fo1} \dot{\nabla}_{\pp_{i}} l^{fo1} \\
		\widehat{\lamlam}_{i+n} = \widehat{\lamlam}_{i}+ k_{p}^{tr2} \nabla_{\dot{\pp}_{i}} l^{tr2}  + k_{p}^{fo2} \nabla_{\dot{\pp}_{i}} l^{fo2}
	\end{cases},
\end{equation}
since $(\widehat{\lamlam}_{i}, \widehat{\lamlam}_{i+n}) \approx (\lamlam_{i}, \lamlam_{i+n})$ represents 
an approximated gradient dynamics for the Lagrangian multipliers whenever the control gains can be selected to satisfy\footnote{Up to some maximum tolerance established by design.} 
\begin{equation}\label{eq:diffeqlamapprox}
	\forall i=1,\ldots,n:~~\begin{cases}
		- \nabla_{\pp_{i}} l^{tr1} \approx k_{d}^{tr1} \ddot{\nabla}_{\pp_{i}} l^{tr1} + k_{p}^{tr1} \dot{\nabla}_{\pp_{i}} l^{tr1}    \\
		- \nabla_{\pp_{i}} l^{fo1} \approx k_{d}^{fo1} \ddot{\nabla}_{\pp_{i}} l^{fo1} +k_{p}^{fo1} \dot{\nabla}_{\pp_{i}} l^{fo1}   \\
		-\nabla_{\dot{\pp}_{i}} l^{tr2} \approx k_{p}^{tr2} \dot{\nabla}_{\dot{\pp}_{i}} l^{tr2}  \\
		-\nabla_{\dot{\pp}_{i}} l^{fo2} \approx k_{p}^{fo2} \dot{\nabla}_{\dot{\pp}_{i}} l^{fo2} 
	\end{cases}.
\end{equation} 
It is also worth to notice that whenever $\sigma^{\prime}_{d_{ij}} \geq 0$ for all $(i,j)\in \mathcal{E}$, input $\widehat{\uu}$ not only approximates $\uu^{\star}$ but is also \textit{practically optimal}, as \eqref{eq:PMPsuff} is satisfied. For this reason, drawing inspiration from the setup illustrated in Sec. \ref{subsec:PRONTOsolvesOIFT},
term $\dot{\nabla}_{\mathbf{p}_{i}}^{safe} l^{fo1} = -\sum_{j\in \mathcal{N}_{i}} \mathcal{H}^{safe}_{\pp_{i}\pp_{j}} l^{fo1} \dot{\ee}_{ij}$ is used in $\widehat{\uu}$ instead of $\dot{\nabla}_{\mathbf{p}_{i}} l^{fo1} = -\sum_{j\in \mathcal{N}_{i}} \mathcal{H}_{\pp_{i}\pp_{j}} l^{fo1} \dot{\ee}_{ij}$. However, the presence of negative $\sigma_{d_{ij}}^{\prime}$ in the term $\nabla_{\pp_{i}}l^{fo1}$ cannot be altered with this heuristics because the latter is part of $\nabla_{\xx} l^{st}$, i.e. it appears directly in the co-state equation \eqref{eq:PMPcostates}. Indeed, $\nabla_{\pp_{i}}l^{fo1}$ is vital for steering the formation positions when repulsive behaviors are required to come into play.
Then, the design criterion $(ii)$ is obeyed, since the structure of $\uu_{i}$ is distributed by construction, i.e. node $i$ uses the information of nodes $j \in \overline{\mathcal{N}}_{i} = \mathcal{N}_{i} \cup \{i\}$ only. In particular, the second-order local estimator in Cor. \ref{lem:secordest} is exploited by each agent to reconstruct the centroid information of the system. 
Moreover, the design criterion $(iii)$ is observed, since $\uu_{i}$ is a function of the current state $\xx(t)$ only and does not require prior computations, e.g. trajectory optimization techniques are not needed.

Finally, the following couple of remarks provide more details on both the theoretical and practical requirements regarding the adoption of \eqref{eq:distrOIFTsol}.

		\begin{rmk}
				Out of their context, equations in \eqref{eq:diffeqlamapprox} can be interpreted as an asymptotically stable dynamics, whenever the constant gains $k_{p}^{tr1}$, $k_{p}^{fo1}$, $k_{d}^{tr1}$, $k_{d}^{fo1}$, $k_{p}^{tr2}$, $k_{p}^{fo2}$ are positive. On the other hand, it is worth to recall that, in this framework, the evolution over time of each variable $(\xx^{\star},\lamlam^{\star},\uu^{\star})$ depends primarily on $\nabla_{\xx} l^{st}(\xx^{\star})$, since $\xx^{\star}$ linearly depends on $\uu^{\star}$ that, in turn, linearly depends on $\lamlam^{\star}$, whose behavior is dictated by structure of $\nabla_{\xx} l^{st}(\xx)$. With such a perspective, the co-state approximation $(\widehat{\lamlam}_{i}, \widehat{\lamlam}_{i+n}) \approx (\lamlam_{i}, \lamlam_{i+n})$ can be satisfied in any scenario, implying that \eqref{eq:lamdynest} is a valid method to approximate the Lagrangian multipliers for large values of $t$ if $m(\xx(T))$ is selected as in \eqref{eq:final_cost_m}. 
    To some extent, the backward integration that emerges from the co-state computation in \eqref{eq:PMPcostates} boils down to the adjustment gains $k_{p}^{tr1},k_{p}^{fo1},k_{d}^{tr1},k_{d}^{fo1},k_{p}^{tr2},k_{p}^{fo2}$: this fact is crucial, since the need for the computation of $2N$ scalar Lagrangian multipliers depending on future time instants is downsized to the choice of a finite number of positive constants.
		\end{rmk}

\begin{rmk}
As far as the distributed online formation tracking control law \eqref{eq:distrOIFTsol} is adopted,
there is no possible means to ensure that all the agents be aware of each mutual position.
For this reason, collisions (especially between non-neighbors) may occur. 
To avoid this kind of issue, collision avoidance protocols taking into account the complete topology w.r.t. the underlying graph should be implemented (see \cite{yang2019tracking}).
Nonetheless, such an approach may be too restrictive; thus, another possible solution is that of equipping each agent with a proximity sensor to guarantee a safety distance radius.
This, in turn, can be put into practice by resorting to repulsive potentials such as that in the first row of \eqref{eq:potential}, e.g. by triggering a repulsive action that is dominant w.r.t. the terms in \eqref{eq:distrOIFTsol} as soon as safety distances are not satisfied. 
\end{rmk}

	\subsubsection{Stability analysis}
	In order to discuss the convergence properties of the control law devised in Ssec. \ref{sec:distr_law} we refer to \cite{SunAndreson2017rigid}. 
	To begin, let us consider the formation terms in \eqref{eq:distrOIFTsol} (marked with ``$fo$'' superscripts): these can be thought as part of a gradient dynamics designed along the same guidelines explored in many research works related to rigid formation control. For this reason, it can be formally proven that the formation stabilization of the given second-order system \eqref{eq:dyn_sys} occurs provided that the underlying network is undirected and infinitesimally rigid (see also \cite{ahn2020book} for more details). Under the latter conditions, it is indeed well known that certain second-order MASs achieve flocking behaviors with both velocity consensus and shape stabilization. In the following lines, we shall clarify that this is also the case when the control law devised in Ssec. \ref{sec:distr_law} is applied. In particular, it is worth showing that the formation terms in \eqref{eq:distrOIFTsol} can be rewritten as a positively-weighted version of equation (12) in \cite{SunAndreson2017rigid}:
	\begin{align}\label{eq:distrOIFTsolformation}
		\widehat{\uu}_{i}^{fo} &= -\mathbf{R}_{i}^{-1}(   k_{p}^{fo1} \nabla_{\pp_{i}} l^{fo1}+
		k_{d}^{fo1} \dot{\nabla}_{\pp_{i}}^{safe} l^{fo1}  + k_{p}^{fo2} \nabla_{\dot{\pp}_{i}} l^{fo2})  \\
		&= -\sum\limits_{j\in \mathcal{N}_i} \mathbf{V}_{ij} (\widehat{\dot{\pp}}_{i}-\widehat{\dot{\pp}}_{j})  - (k_{p}^{fo1}k_{F}\mathbf{R}_{i}^{-1}) \sum\limits_{j\in \mathcal{N}_i} \sigma_{d_{ij}}^{\prime} \!\!\left(\left\|\widehat{\pp}_{i}-\widehat{\pp}_{j}\right\|^{2}_{2}\right)  (\widehat{\pp}_{i}-\widehat{\pp}_{j}) , \nonumber
	\end{align}
	where the weighting factors are expressed by quantities $(k_{p}^{fo1}k_{F}\mathbf{R}_{i}^{-1}) \succ 0$ and
	\begin{equation*}
		\mathbf{V}_{ij} =  -\mathbf{R}_{i}^{-1} (k_{p}^{fo2}\mathcal{H}_{\dot{\pp}_{i},\dot{\pp}_{j}} l^{fo2} + k_{d}^{fo1} \mathcal{H}^{safe}_{\pp_{i},\pp_{j}} l^{fo1}(\widehat{\pp})) \succ 0, \quad \forall \widehat{\pp} \in \mathbb{R}^{N}.
	\end{equation*}
	We then recall that the centroid velocity is a constant of motion for \eqref{eq:distrOIFTsolformation} (see Lemma 3.1 in \cite{SunAndreson2017rigid}). Hence, the formation dynamics \eqref{eq:distrOIFTsolformation} can be completely decoupled from the remaining tracking terms (marked with ``$tr$'' superscript) while analyzing \eqref{eq:distrOIFTsol}. In fact, by setting $\bar{\Cc} = n^{-1} \begin{bmatrix}
		\Ii_{M} & \cdots & \Ii_{M}
	\end{bmatrix} \in \mathbb{R}^{M\times N}$ and $\Qq = \Diag(\Qq_{c},\Qq_{\dot{c}})$, it can be appreciated that the term
	\begin{align}\label{eq:distrOIFTsoltracking}
		\widehat{\uu}_{i}^{tr} &= -\mathbf{R}_{i}^{-1}(k_{p}^{tr1} \nabla_{\pp_{i}} l^{tr1}+k_{d}^{tr1} \dot{\nabla}_{\pp_{i}} l^{tr1} + k_{p}^{tr2} \nabla_{\dot{\pp}_{i}} l^{tr2}) , \\
		&= - \mathbf{R}_{i}^{-1} \bar{\Cc}^{\top} \left[ k_{p}^{tr1} \Qq_{c} (\bar{\Cc} \widehat{\pp} - \pp_{c,des}) +  (k_{d}^{tr1} \Qq_{c}+k_{p}^{tr2}\Qq_{\dot{c}}) (\bar{\Cc} \widehat{\dot{\pp}} - \dot{\pp}_{c,des}) \right] \nonumber
	\end{align}
	only affects the tracking\footnote{\rev{It can be also observed that action \eqref{eq:distrOIFTsoltracking} minimizes the cost in \eqref{eq:track_inst_cost}, as $l^{tr}$ is quadratic in $\xx(t)$.}} of both the desired centroid position and velocity. Consequently, the tracking term in \eqref{eq:distrOIFTsoltracking} does not influence the equilibria attained by applying \eqref{eq:distrOIFTsolformation}. As a result, the same flocking behavior and shape stabilization of the original protocol (12) in \cite{SunAndreson2017rigid} are preserved upon the linear combination $\widehat{\uu}_{i} = \widehat{\uu}_{i}^{tr}+\widehat{\uu}_{i}^{fo}$.\\
	Additionally, it is also worth to notice that, since $\widehat{\uu}_{i}$ is a gradient-based law, it is stabilizing for both formation and tracking. Whenever the final trait of the desired path is required to be covered at a constant velocity, one has $\widehat{\uu}(T) \rightarrow \uups $ for large enough $T$, where $\uups$ is some (small) constant due to the fact that a second-order integrator controlled through proportional-derivative feedback reaches zero steady-state error. 
	The expression in \eqref{eq:ineqerrbase2} for the local second-order estimator presented in Cor. \ref{lem:secordest} can be thus used to roughly bound the steady-state estimation error at time $t=T$.

\subsubsection{Heuristics to limit the energy consumption}
To conclude the discussion, we illustrate a simple method that can be considered to limit the input energy. The adoption of a saturated version $\bar{\uu}$ of input $\widehat{\uu}$ can be indeed employed to refine approximation errors made by \eqref{eq:lamdynest}, since high values for the feedback gains imply high values for the instantaneous input energy $l^{in}(\widehat{\uu})$. In formulas, for all individual scalar components $j = 1,\ldots,N$, assign
\begin{equation}\label{eq:saturation_heuristics}
	\bar{u}_{j} = \begin{cases}
		\widehat{u}_{j}, \quad &\text{if } \left|\widehat{u}_{j}\right| \leq U; \\
		\mathrm{sign}(\widehat{u}_{j})U, \quad &\text{otherwise.}
	\end{cases}
\end{equation}
where $\mathrm{sign}$ is the classic sign function (defined as $0$ at $0$) and $U>0$ is a suitable saturation constant inversely proportional to each feedback gain. \\

\section{Numerical results} \label{sec:numerical_simulations}
In this section, we focus on a singular case study where desired distances $d_{ij}$ are imposed such that the formation assumes a cubic shape in a space of dimension $M=3$. Precisely, the desired cube has side equal to $d = 5~\si{\meter}$ and each of the $n=8$ agents have a neighborhood with degree $|\mathcal{N}_{i}| = 5$, $i=1,\ldots, n$. The following adjacency matrix summarizes the distance constraints to be satisfied (the symbol $*$ is used to indicate the absence of a link):
\begin{equation*}
		\begin{bmatrix}
			0 & d & * & d & d & \sqrt{2}d & * & \sqrt{2}d \\
			d & 0 & d & \sqrt{2}d & * & d & * & \sqrt{3}d \\
			* & d & 0 & d & \sqrt{3}d & \sqrt{2}d & d & * \\
			d & \sqrt{2}d & d & 0 & * & * & \sqrt{2}d & d \\
			d & * & \sqrt{3}d & * & 0 & d & \sqrt{2}d & d \\
			\sqrt{2}d & d & \sqrt{2}d & * & d & 0 & d & * \\
			* & * & d & \sqrt{2}d & \sqrt{2}d & d & 0 & d \\
			\sqrt{2}d & \sqrt{3}d & * & d & d & * & d & 0
		\end{bmatrix}
\end{equation*}
This setup allows us to present an example where the formation to be achieved is infinitesimally rigid \cite{jordan2016ii}. 
Thus, the convergence for the whole system towards a unique stable equilibrium is ensured by such topology and geometry. Also, the group of agents have to track 
the curve $\Gamma(t) = \mathcal{R}_{z}(\frac{\pi}{4})\mathcal{R}_{y}(-\frac{\pi}{4})\Gamma_{0}(t)$, where $\mathcal{R}_{z}$, $\mathcal{R}_{y}$ denote the three-dimensional rotation matrices about the $y$ and $z$ axes, respectively, and
\begin{equation}
	\Gamma_{0}(t) :\quad \begin{cases}
		x(t) = 2 t\\
		y(t) = 10 \tanh(10(t-T/2)) \\
		z(t) = 0
	\end{cases}, \quad t \in [0,T]. 
\end{equation}
The tracking velocity is set to be equal to $\dot{\Gamma}(t)$.
The initial state $\xx_{0} = \vc{\pp_{0},\dot{\pp}_{0}}$ of the system is randomly assigned as follows:
\begin{align*}
	\mathbf{p}_{0} &= \vc{\begin{bmatrix}
			-6 & -9 & 6 & 15 & -3 & 6 & -3 & 15 \\
			3 & -3 & -6 & 3 & -18 & 6 & -15 & 6 \\
			24 & 3 & 15 & 15 & 6 & 6 & -3 & 9
	\end{bmatrix}} 
	\si{\meter};\\
	\dot{\mathbf{p}}_{0} &= \vc{\begin{bmatrix}
			10 & 10  & 0 & 15 & 0 & 5 & 0 & -10 \\
			1 & 0 & 0 & -5 & 0 & 5 & 0 & -25 \\
			0 & 0 & 0 & 5 & 5 & 5 & 0 & 10
	\end{bmatrix}} \si{\meter\second^{-1}}.
\end{align*}

We select the input weighting matrix in \eqref{eq:inst_energy} as $\mathbf{R} = \mathbf{I}_{N}~\si{\meter^{-2} \second^{4}}$. The integration time $T=20 ~\si{\second}$ is kept fixed for all simulations, as well as the structure of the output weighting matrices $\mathbf{Q}_{i}= \mathrm{Diag}(q_{p} \mathbf{I}_{M}, q_{d}\mathbf{I}_{M})$, $i = 1,\ldots, n$,
for \eqref{eq:track_inst_cost}, where $q_{p} = 1.25~\si{\meter}^{-2}$, $q_{d} = 0.125~\si{\meter}^{-2}\si{\second}^{2}$ are chosen.
Moreover, we set, for all edges $(i,j)\in\mathcal{E}$, the following parameters: $k_{r_{ij}} = 250$, $k_{a_{ij}} = 100$, $\beta_{ij}=3$, $\alpha_{ij}=0.5$, in \eqref{eq:potential}, $k_{F} = 2$, in \eqref{eq:form_inst_cost} and $k_{A} = 0.25$, $\Thth_{ij} = \Ii_{M}~\si{\meter}^{-2}\si{\second}^{2}$, in \eqref{eq:align_inst_cost}. In each numerical simulation presented, we have decided to stop the execution of PRONTO 
by taking two actions: either the algorithm stops when the iteration $k$ exceeds the preset number of iterations $MaxIter = 80$
or the search direction becomes sufficiently flat. 

Constant gains for the PD controller in \eqref{eq:PDcontroller} are selected leveraging the second-order linear dynamics of \eqref{eq:dyn_sys}, as
\begin{equation*}
	\begin{cases}
		\omega_{n} = 3 ~\si{\radian\per\second}\\
		\xi = 0.7 
	\end{cases} \Rightarrow ~ \begin{cases}
		k_{p} = \omega_{n}^{2} \\
		k_{d} = 2\xi\omega_{n}
	\end{cases}.
\end{equation*}
Whereas, for the online distributed feedback controller in \eqref{eq:distrOIFTsol} gains $k_{p}^{tr1} = 2.4~\si{\second}^{2}$, 
$k_{p}^{fo1} = 1.3~\si{\second}^{2}$, $k_{d}^{tr1} = 1.2~\si{\second}^{3}$, 
$k_{d}^{fo1} = 1~\si{\second}^{3}$, $k_{p}^{tr2} = 12~\si{\second}$, 
$k_{p}^{fo2} = 0.3~\si{\second}$ have been tuned by matching the distributed solution with the centralized one provided by PRONTO. Estimator gains in \eqref{eq:estimator2} are set as $k_{py}=180~\si{s}^{-1}$, $k_{dy}=170~\si{s}^{-1}$, such that, omitting the physical dimensionality, condition $k_{py}>k_{dy} >> k_{p}^{tr1},k_{p}^{fo1},k_{d}^{tr1},k_{d}^{fo1},k_{p}^{tr2},k_{p}^{fo2}$ is satisfied to guarantee fast enough estimation dynamics, yet avoiding noisy estimation behaviors. Finally, as suggested in \eqref{eq:saturation_heuristics}, a saturation $|\widehat{u}_{j}| \leq 50 ~\si{\meter}\si{\second}^{-2}$ has been selected for each control component $\widehat{u}_{j}$ of input $\widehat{\uu}$.

\subsection{Performance of the devised distributed controller}
With the purpose of validating the online distributed feedback controller in \eqref{eq:distrOIFTsol}, we compare the two dynamics obtained by the latter control law and the offline inverse dynamics provided by PRONTO. As an evaluation criterion, we decide to look at the global energy spent by the input and the settling time of the system dynamics. Specifically, we account for the global input energy $l^{in}(\cdot)$ and the average input energy $\bar{l}^{in}(\cdot)$. This quantity is defined as the weighted mean $\bar{l}^{in}(\cdot) = l^{in}(\cdot)/\left\|\mathbf{R}\right\|_{F}$, where $\left\|\mathbf{R}\right\|^{2}_{F} = \sum_{i=1}^{n} \sum_{j=1}^{n} [\mathbf{R}]_{ij}^{2}$ has to be considered the dimensionless version of the Frobenius norm of $\mathbf{R}$.
Furthermore, let us introduce function $l^{tf}:\mathbb{R} \rightarrow [-1,1]$, defined 
as
\begin{equation}\label{eq:lst}
	l^{tf}(t) = \begin{cases}
		\dfrac{l^{fo}(t)-l^{tr}(t)}{l^{fo}(t)+l^{tr}(t)},  \quad &\text{if } l^{fo}(t)+l^{tr}(t) > 0;\\
		0, \quad &\text{otherwise}.
	\end{cases}
\end{equation}

\begin{figure}[]
	\centering
	\subfigure[PRONTO: Position trajectories]{\includegraphics[width=0.495\columnwidth, trim=180 0 200 60, clip]{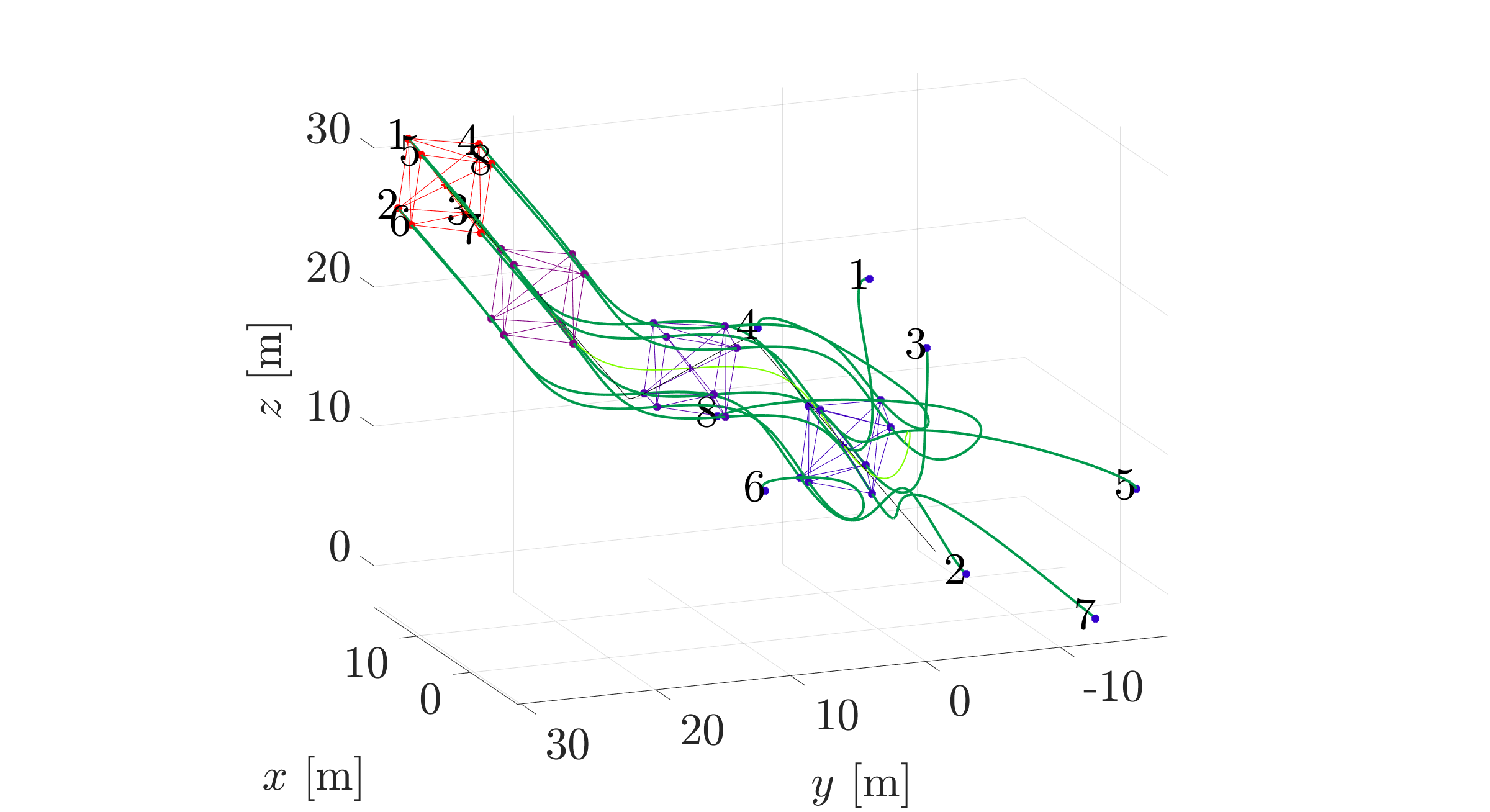}\label{fig:perftrajPRONTO}}\hspace{0cm}
	\subfigure[Distributed: Position trajectories]{\includegraphics[width=0.495\columnwidth, trim=180 0 200 60, clip]{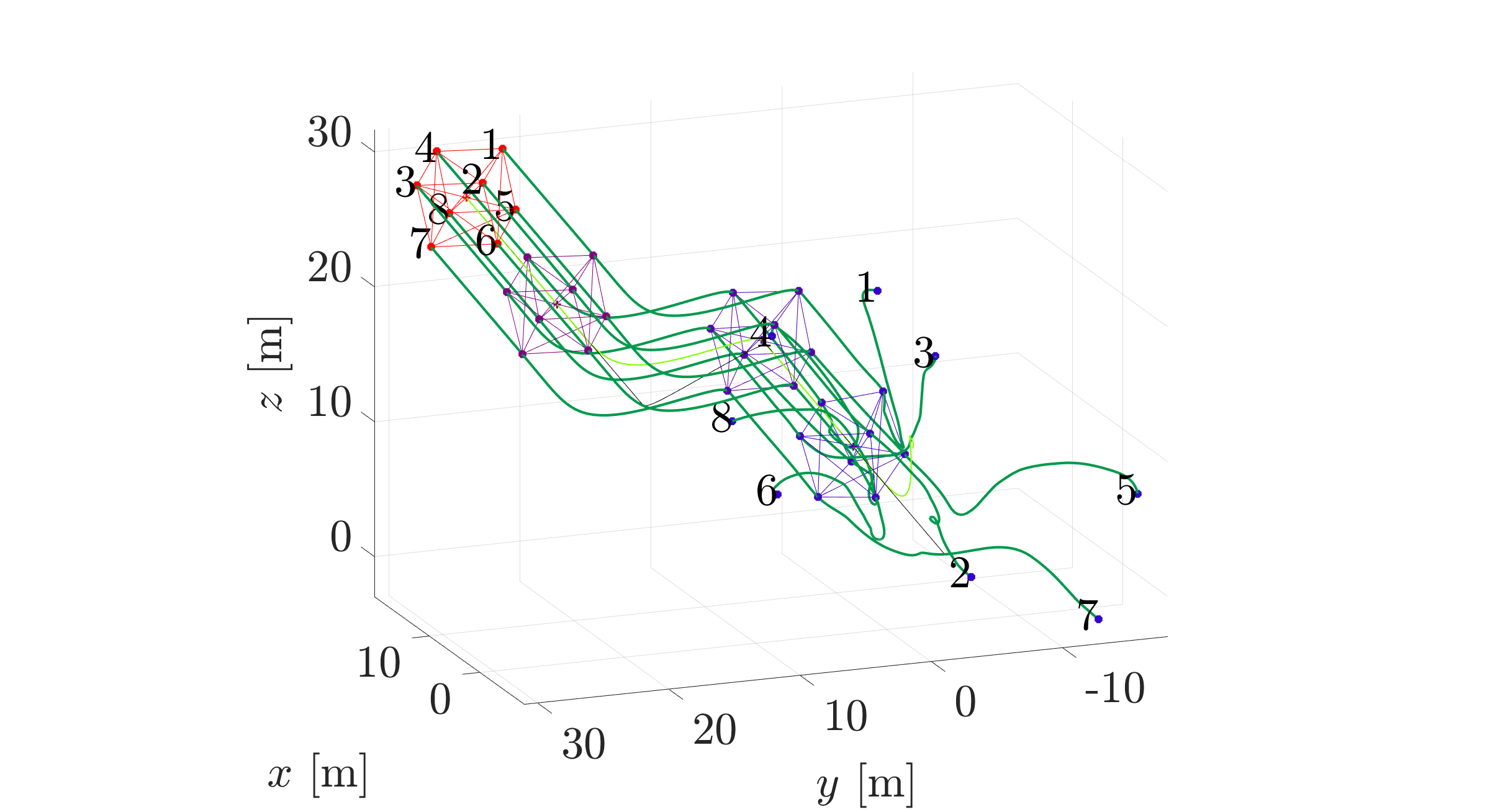}\label{fig:perftrajdistr}}\vspace{-0cm}\\
	\subfigure[PRONTO: Input energy consumption]{\includegraphics[width=0.495\columnwidth, trim= 40 5 85 30, clip]{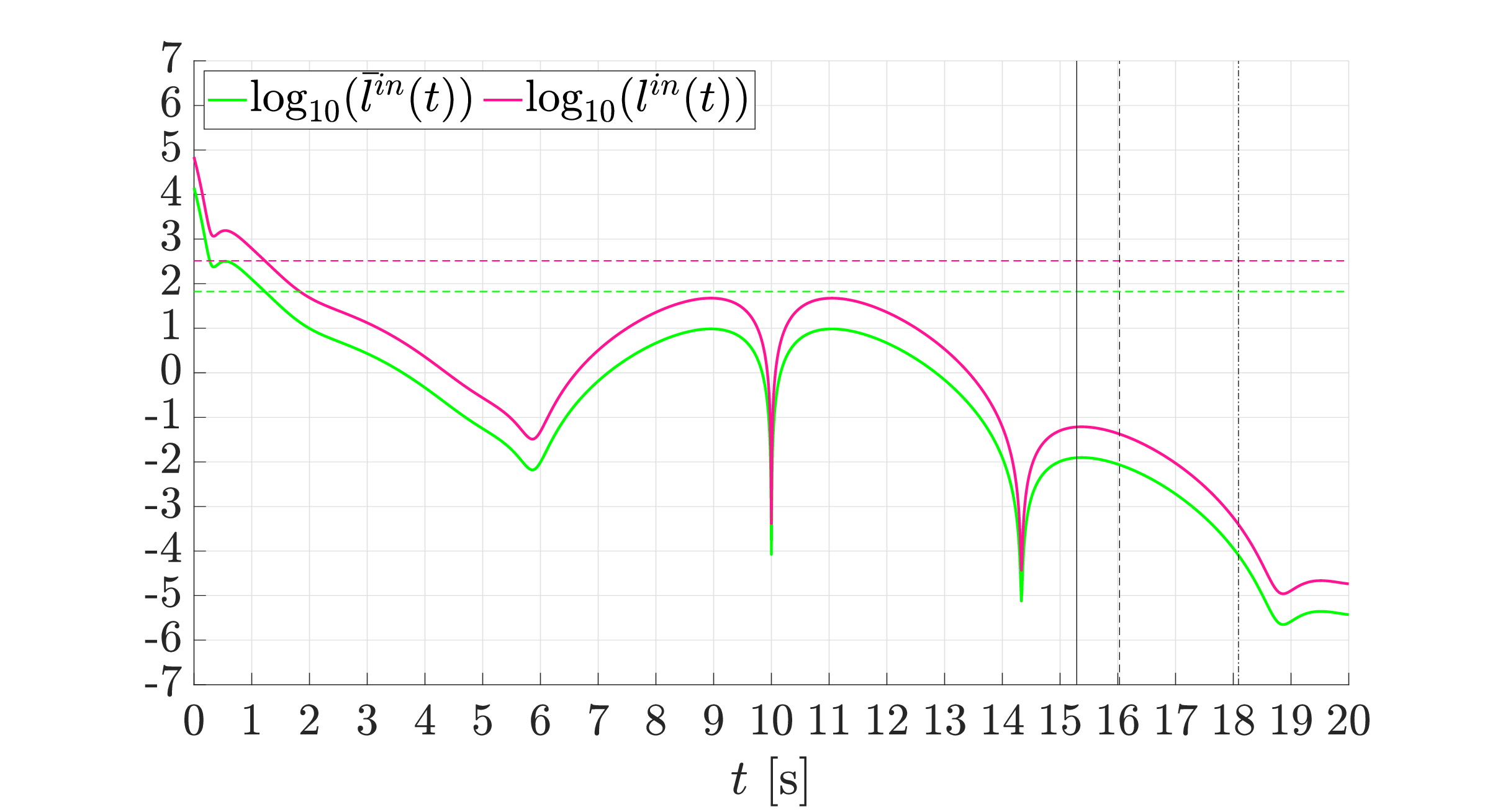}\label{fig:perfenerPRONTO}}\hspace{0cm}
	\subfigure[Distributed: Input energy consumption]{\includegraphics[width=0.495\columnwidth, trim= 40 5 85 30, clip]{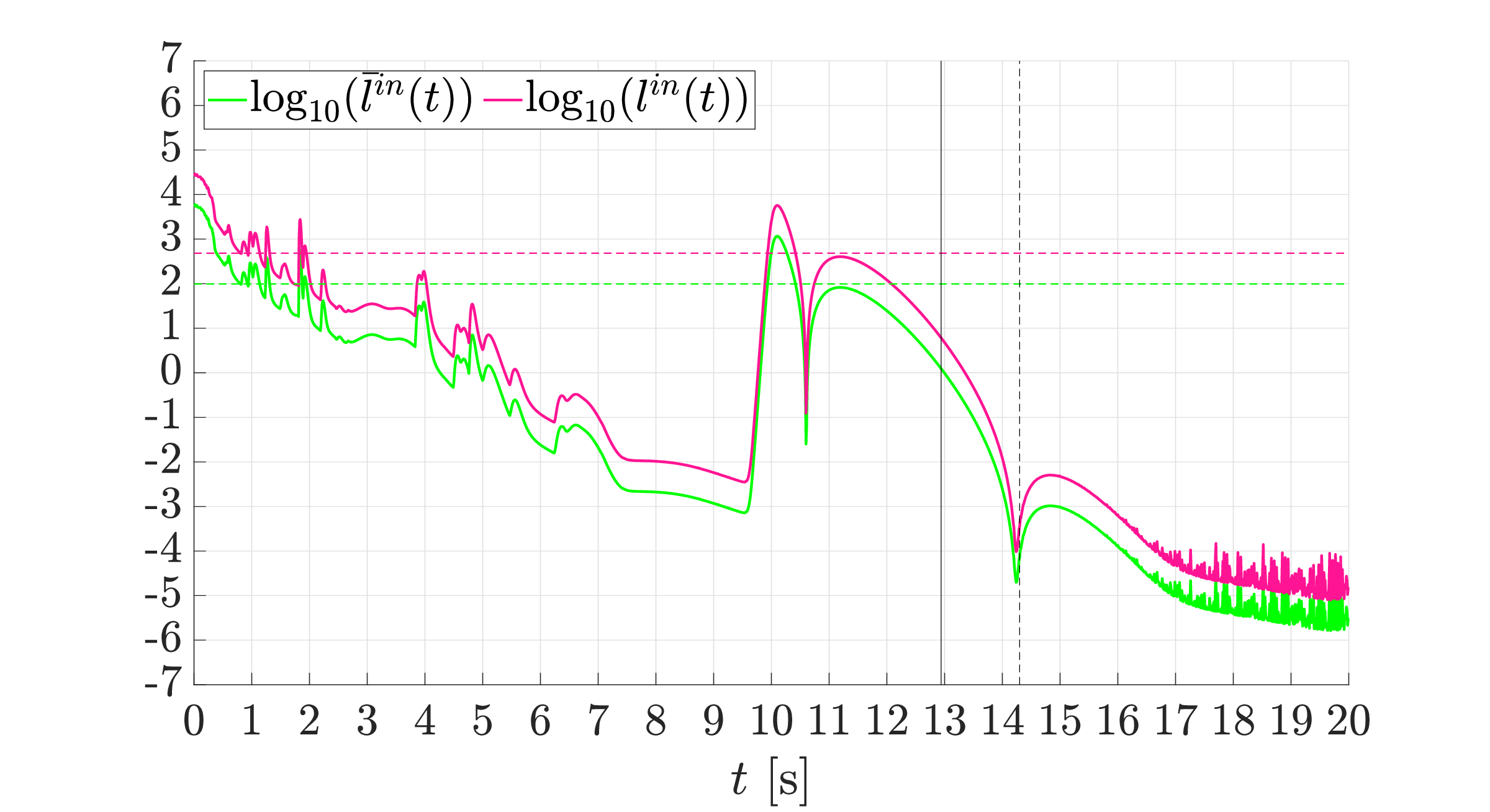}\label{fig:perfenerdistr}}\vspace{-0cm}\\
	\subfigure[PRONTO: Settling time]{\includegraphics[width=0.495\columnwidth, trim= 30 5 85 30, clip]{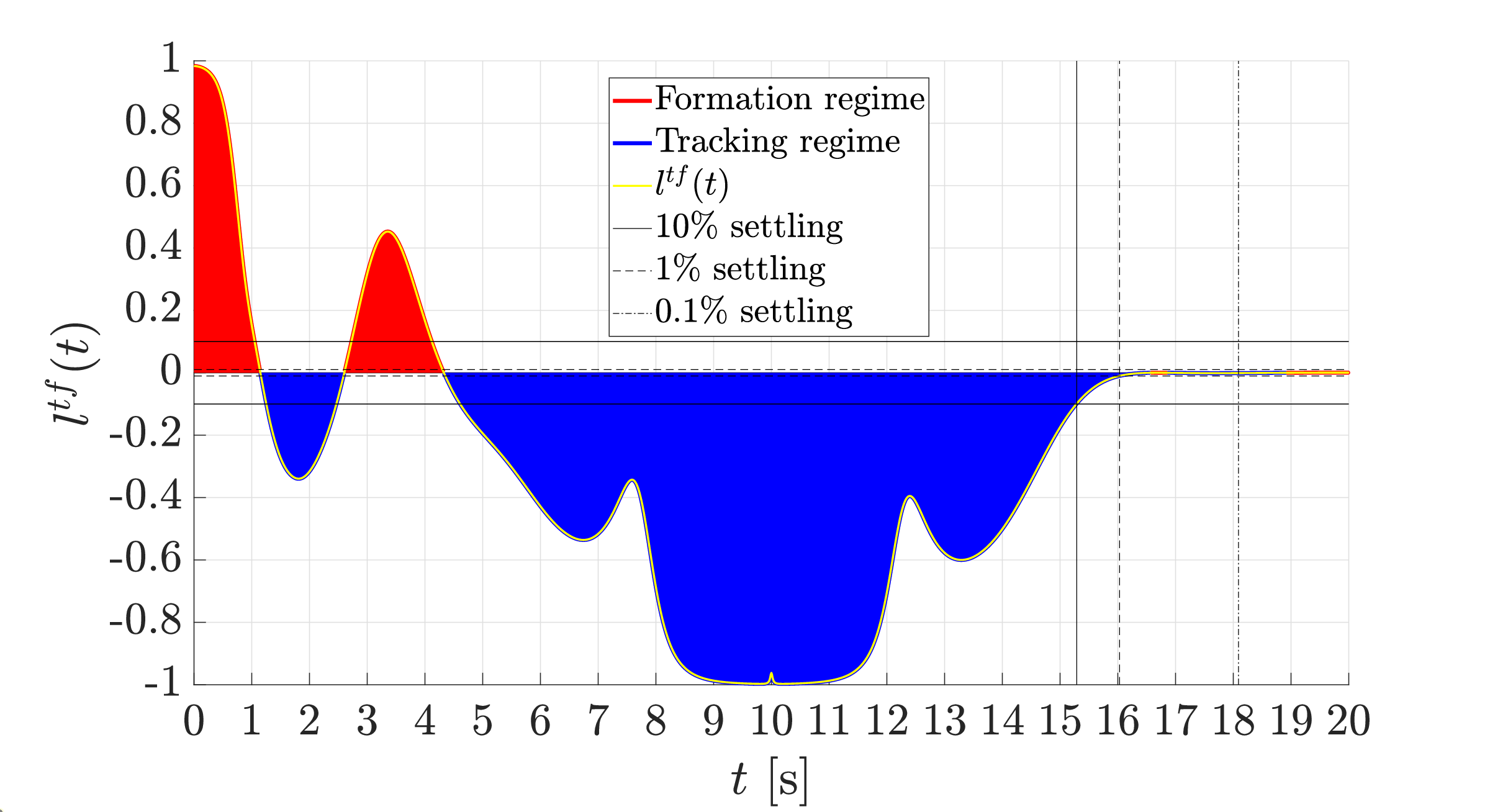}\label{fig:perfsettlPRONTO}}\hspace{0cm}
	\subfigure[Distributed: Settling time]{\includegraphics[width=0.495\columnwidth, trim= 30 5 85 30, clip]{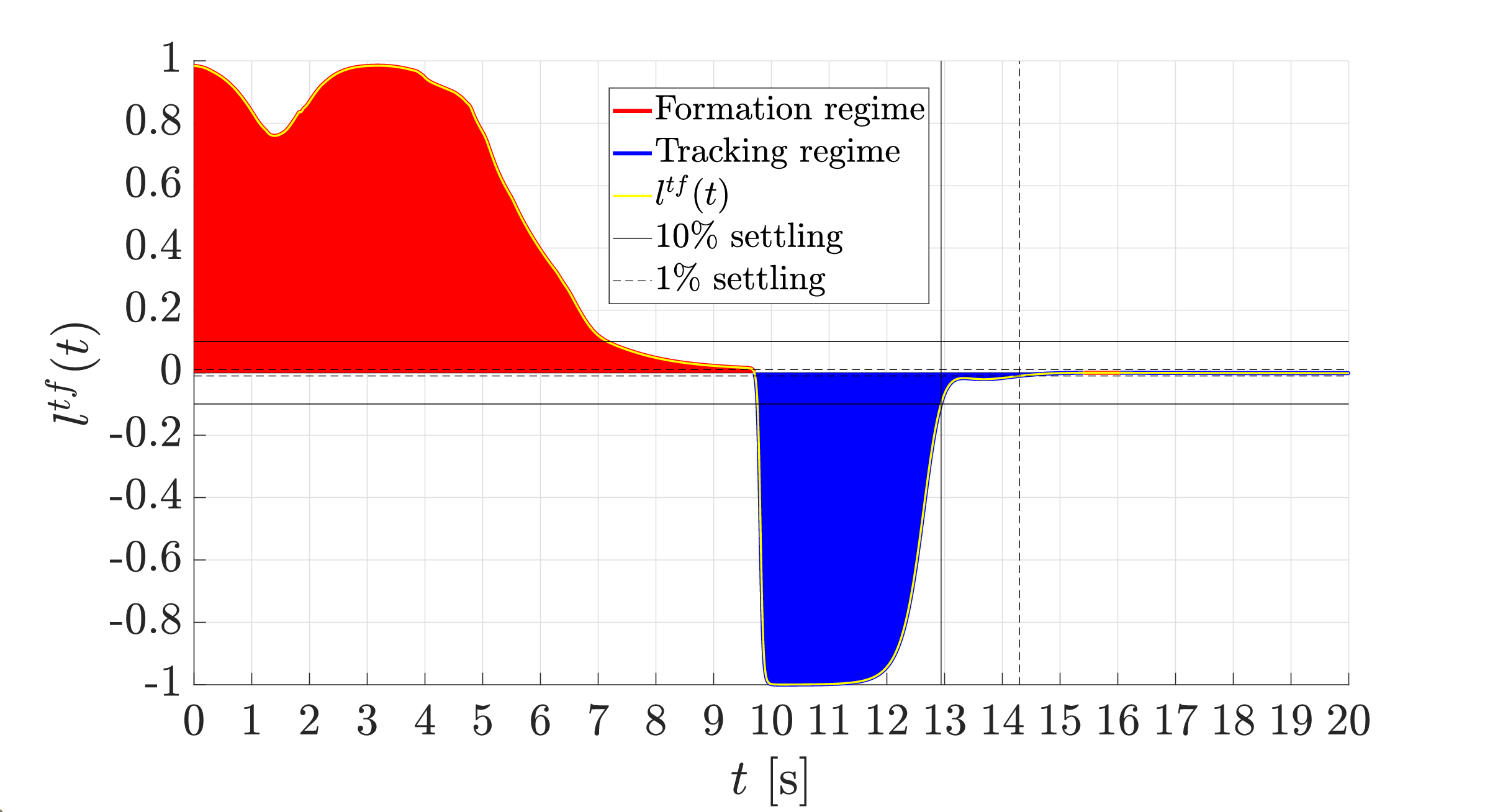}\label{fig:perfsettldistr}}\vspace{-0cm}\\
	\caption{Comparison between the offline solution provided by PRONTO and the system dynamics governed by the online distributed feedback controller. (a)-(b): trajectories of the positions, in dark green; trajectory of the centroid position, in light green; desired straight path, in black; cubic shape formation, depicted progressively with ordered shades blue-indigo-purple-red, beginning with blue. (c)-(d): total input energy, in magenta and average input energy, in green, spent by the system; the dashed lines show their time-averaged values over the interval $[0,T]$. (e)-(f): settling times at $10\%$, $1\%$, $0.1\%$, (i.e. $\delta = 0.1,0.01,0.001$ respectively) depicted by the vertical lines, and established evaluating a quantitative weighted trade-off between formation and tracking instantaneous costs.}
	\label{fig:performancesPRONTOdistr}	
\end{figure}

Selecting $\delta\in[0,1]$, the settling time instant $t_{\delta} \in [0,T]$ for the system is established whenever condition $|l^{tf}(t)| \leq \delta$ holds for all $t \geq t_{\delta}$. 
It is worth to note that function in \eqref{eq:lst} measures a weighted trade-off between centroid-based and relative-based control performances, as it is expected to go to $0$ and remain null over time if a control action is being applied.
In Fig. \ref{fig:performancesPRONTOdistr} the performances of these numerical simulations are illustrated, presenting the inverse dynamics obtained by PRONTO on the left column and the online dynamics governed by the distributed feedback controller on the right column. In Figs. \ref{fig:perftrajPRONTO} and \ref{fig:perftrajdistr}, one can observe the positional trajectories described by the system in a three-dimensional space, noting that each agent contributes eventually to achieve a proper position as a vertex of the desired cubic shape. Moreover, the system's centroid tracks the required curve 
$\boldsymbol{\Gamma}(\cdot)$. 
The most evident difference between the two approaches arises by comparing how the trajectories perform in the chicane at the middle of the desired path. PRONTO provides a solution based on future knowledge; hence, the chicane tracking is planned ahead and is carried out in a symmetric way w.r.t. $t=T/2=10 ~\si{\second}$. Conversely, the control law devised in \eqref{eq:distrOIFTsol} exhibits a causal behavior, stabilizing only after the transient of $\Gamma(t)$ has occurred.
It is also worth to note that the trajectories of the system governed by the distributed control law are less smooth w.r.t. those of PRONTO. This indicates that more effort is needed to steer the agents using the distributed controller \eqref{eq:distrOIFTsol}, especially when the ``$safe$'' heuristic for term $\dot{\nabla}_{\pp} l^{fo1}$ comes into play. Indeed, Fig. \ref{fig:perfenerdistr} depicts a higher energy consumption levels over time w.r.t. to that in Fig. \ref{fig:perfenerPRONTO}, despite the saturation. 
Figs. \ref{fig:perfsettlPRONTO} and \ref{fig:perfsettldistr} again highlight the fact that, with PRONTO, the tracking effort (blue area) spent in the chicane is better ``allocated'' over time and it kicks in before $t=T/2$. Whereas, for the distributed online controller, the tracking effort is maximum at $t=T/2$, and before that instant the formation task is mainly addressed. These plots also show that each settling time instant can be compared among the two approaches with a maximum deviation of about $2~\si{\second}$. 
Remarkably, in this case, the distributed controller reaches the $1\%$ 
settling threshold at $t \simeq 15 < T$, 
even before the trajectory planned by PRONTO does at $t = 16 < T$. 
The explanation of this fact lays on the slightly greater energy spent by the distributed controller. However, the distributed control does not settle at $0.1\%$ because the estimation of a local frame introduces some inaccuracy\footnote{Settling at $0.1\%$ would talk place if exact information on the centroid was available.}.

To summarize, the latter observation tells us that law in \eqref{eq:distrOIFTsol} works as a finite-time practically optimal controller, according to the choice of function $l^{tf}$ as a reference for the settling time measurements.
As a final remark, three important evidences can be highlighted for the law in \eqref{eq:distrOIFTsol}:
\begin{itemize}[]
	\item it allows to govern the state dynamics of the system similarly to what PRONTO algorithm produces as optimal inverse dynamics;
	\item it is practically optimal, since it requires the system to spend slightly more input energy to achieve similar time performances w.r.t. the optimal inverse dynamics;
	\item for infinitesimally (\rev{resp.} globally) rigid geometrical shapes the \rev{local (resp. global)} convergence is attained in a finite time, provided that gains of the distributed PD controller are properly tuned.
\end{itemize}

\subsection{Convergence behavior of the devised distributed controller}
It is immediate to note that two control actions -- relative distance and centroid regulations -- can be analyzed separately in law \eqref{eq:distrOIFTsol}. Indeed, it can be pointed out that the relative distance regulation acts by exploiting the potentials in \eqref{eq:potential}, as distance-based error functions and the second-order formation term \eqref{eq:align_inst_cost}. Such an approach can be seen as a generalized consensus protocol \cite{MesbahiEgerstedt2010} for the dynamics under investigation.

\begin{figure}[th!]
	\centering
	\subfigure[Zero-order position consensus]{\includegraphics[width=0.495\columnwidth, trim= 10 5 85 30, clip]{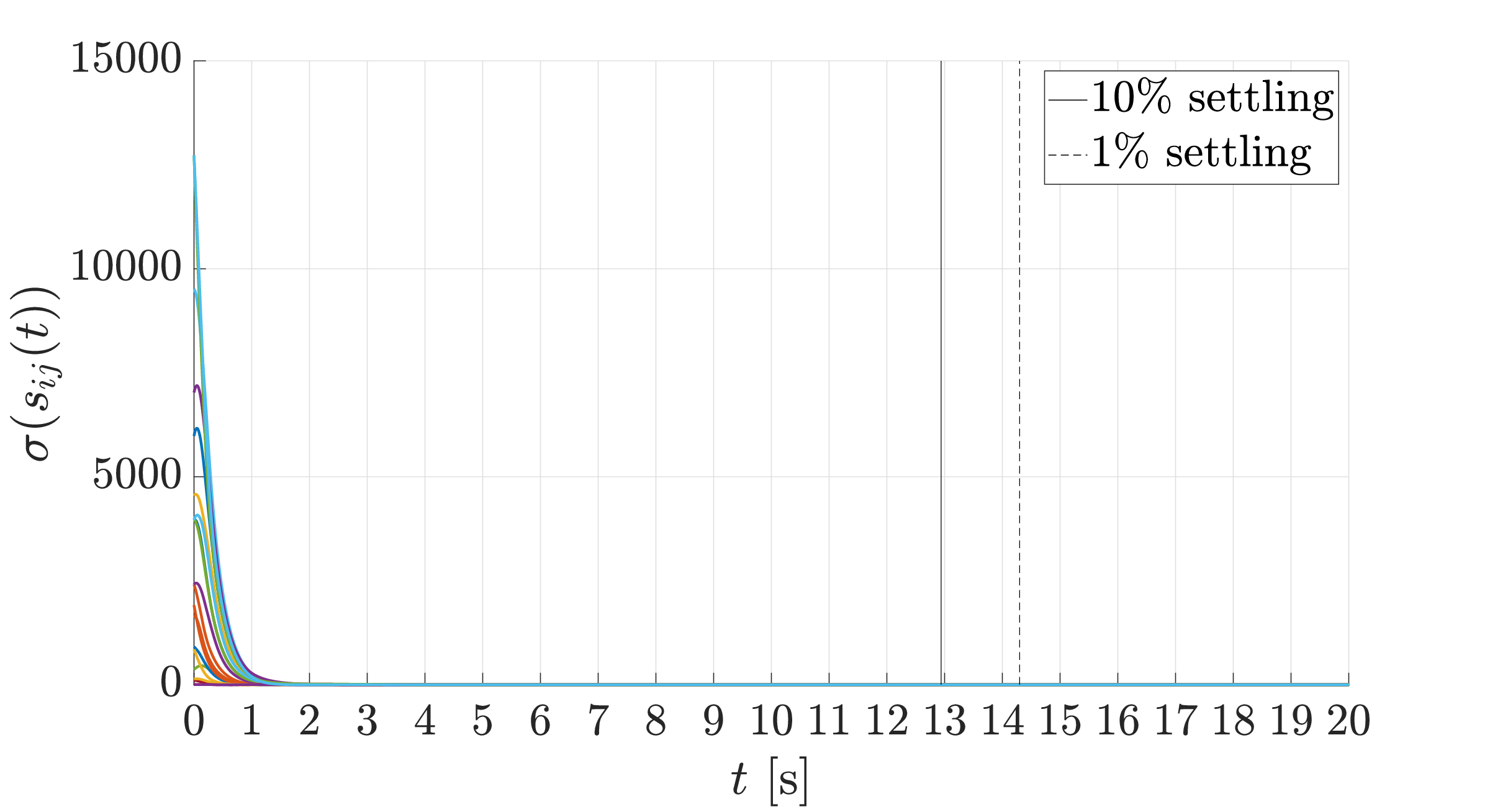}\label{fig:cons0sigma}}\hspace{0cm}
	\subfigure[Velocity consensus]{\includegraphics[width=0.495\columnwidth, trim= 10 5 85 30, clip]{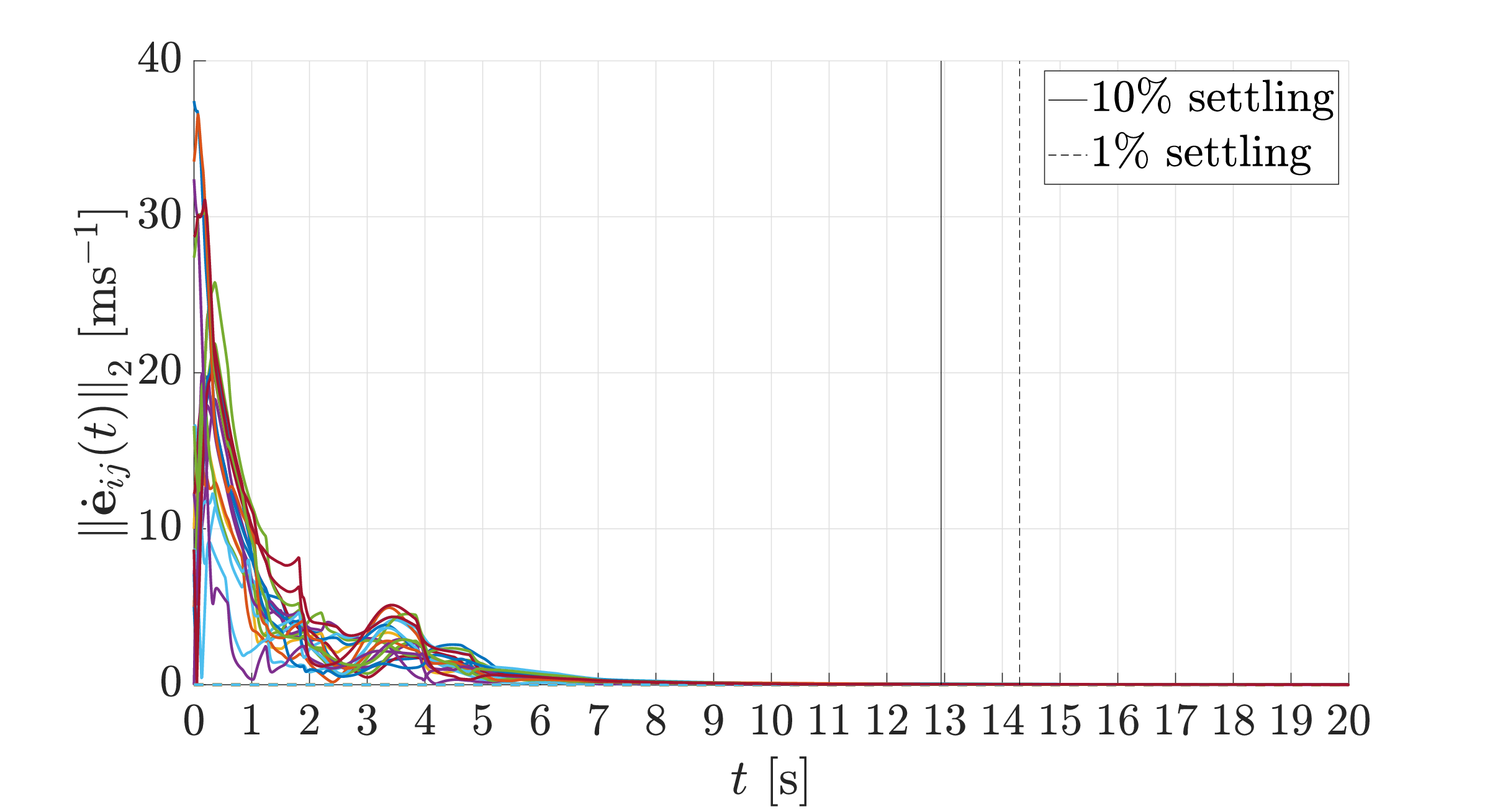}\label{fig:reldistconsensus}}\vspace{-0cm}\\
	\subfigure[First-order position consensus]{\includegraphics[width=0.495\columnwidth, trim= 10 5 85 30, clip]{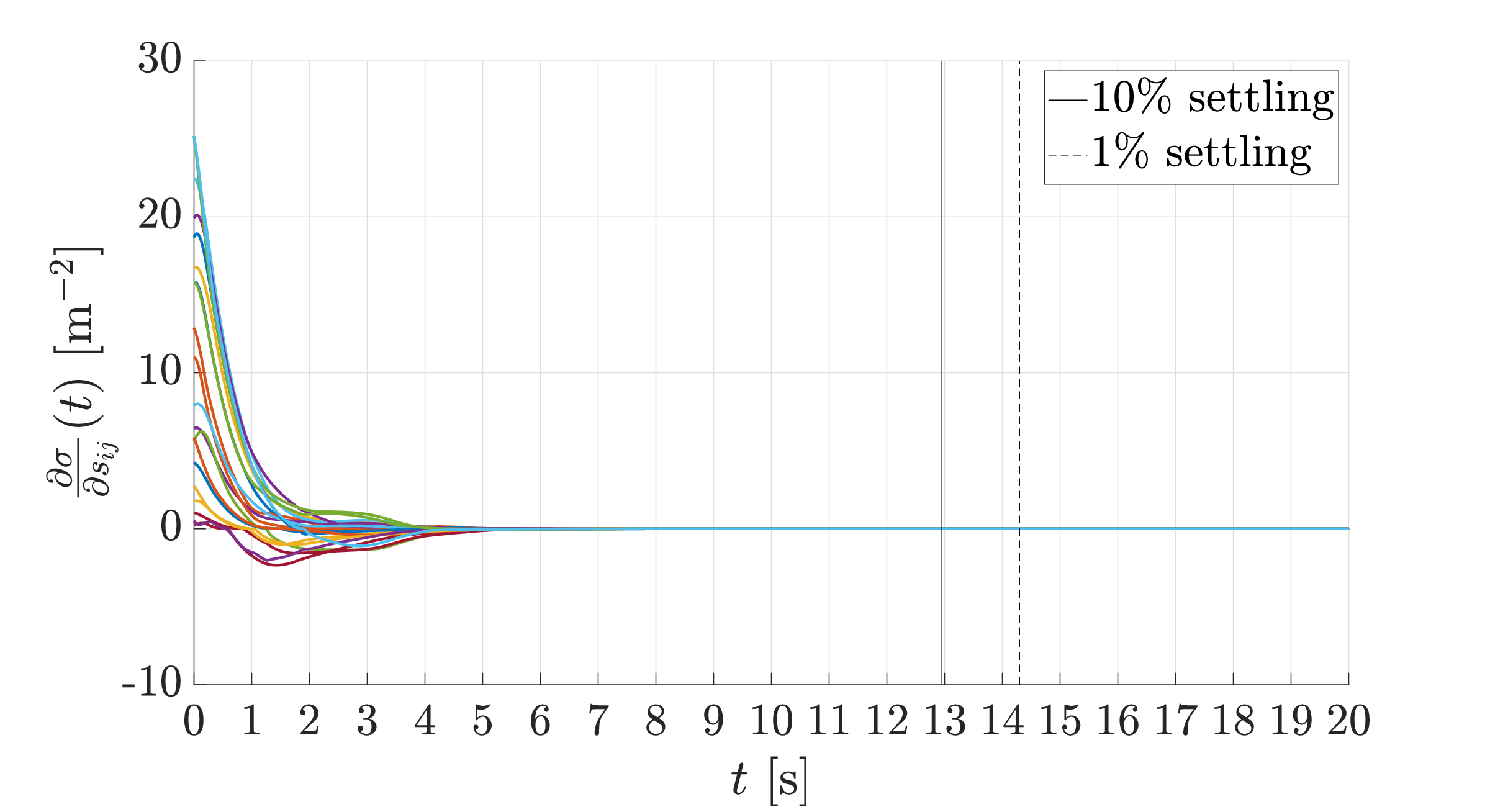}\label{fig:cons1sigma}}\hspace{0cm}
	\subfigure[Centroid position response]{\includegraphics[width=0.495\columnwidth, trim= 10 5 85 30, clip]{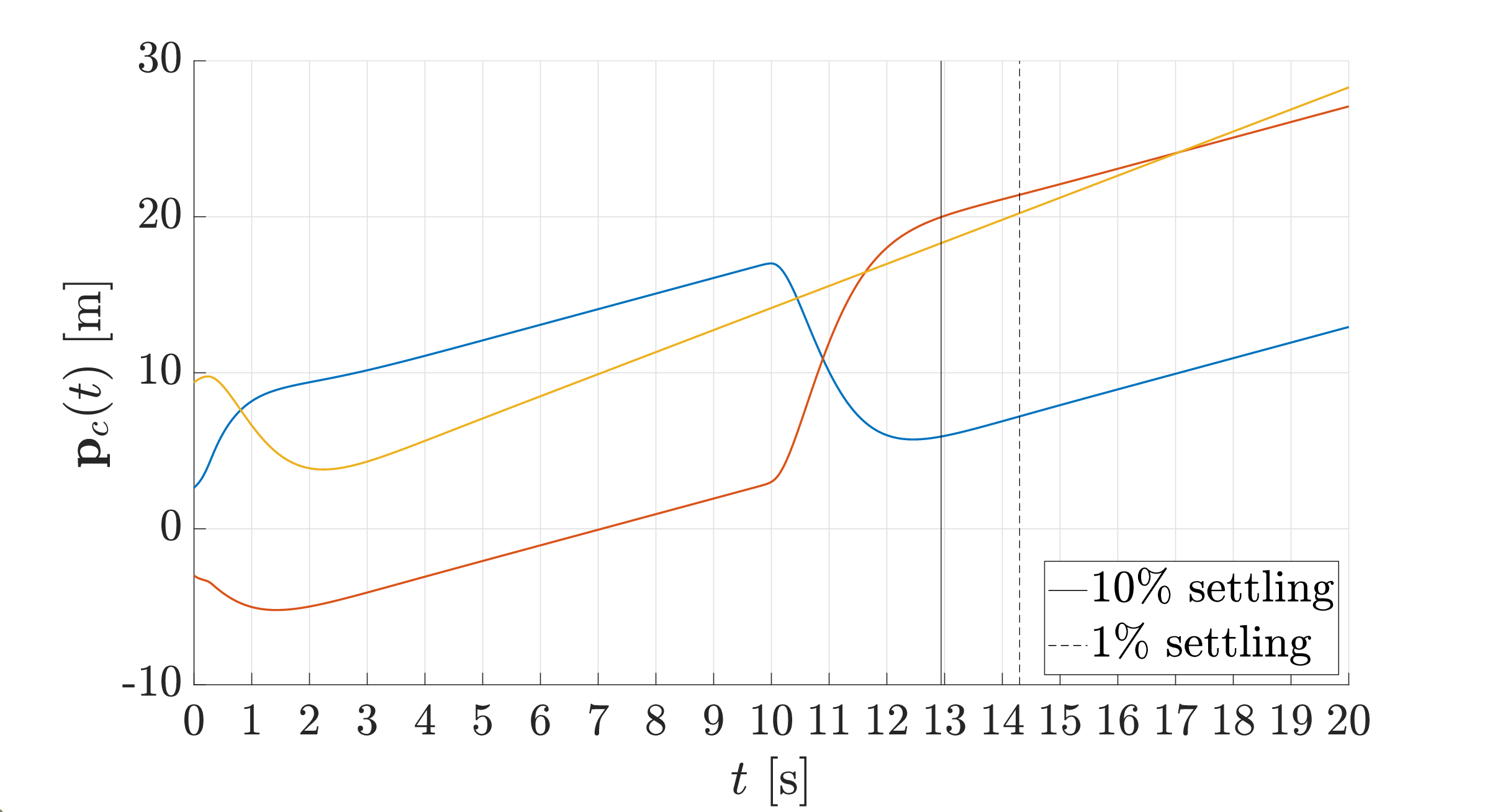}\label{fig:trackpc}}\vspace{-0cm}\\
	\subfigure[Second-order position consensus]{\includegraphics[width=0.495\columnwidth, trim= 10 5 85 30, clip]{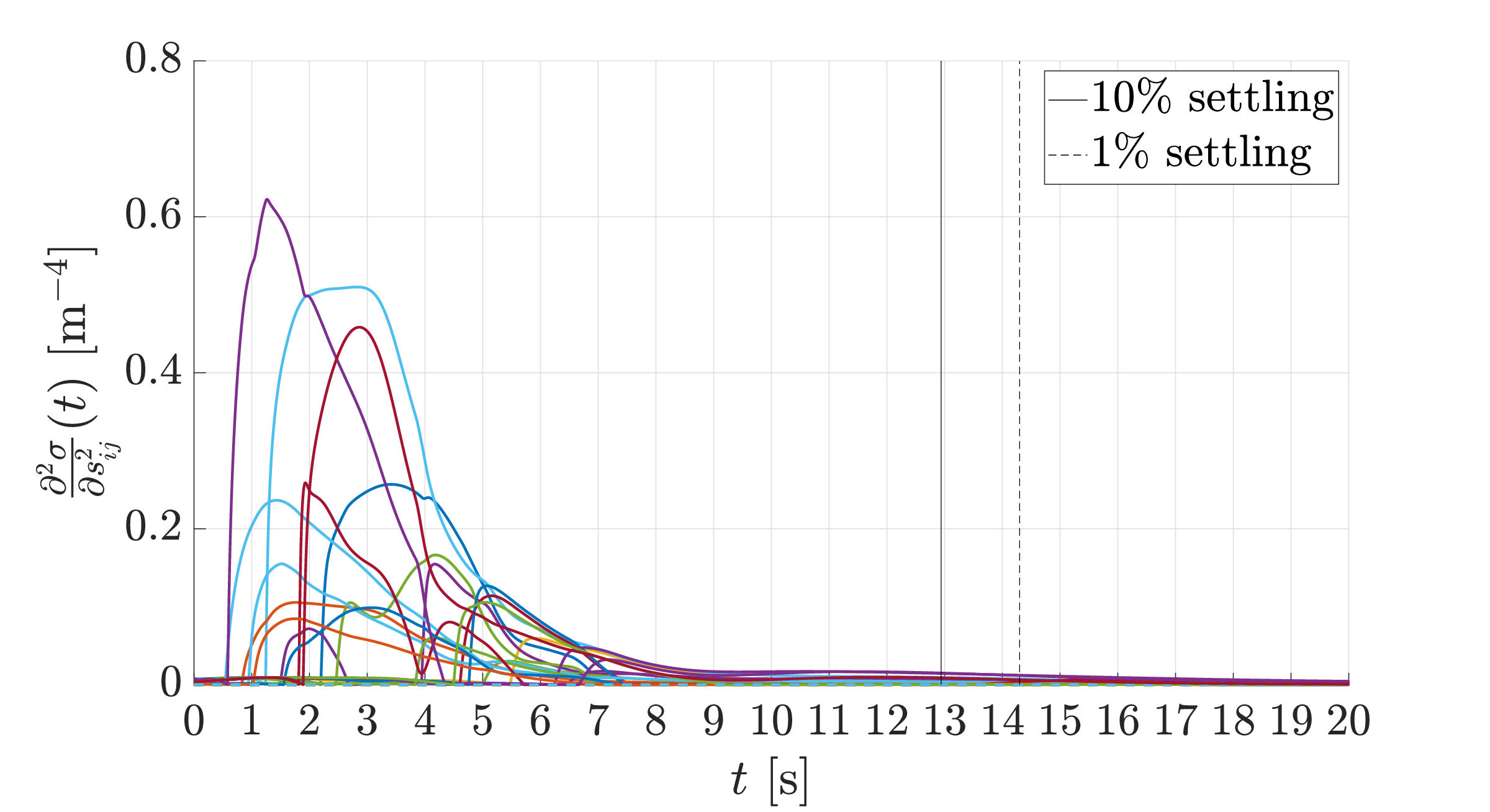}\label{fig:cons2sigma}}\hspace{0cm}
	\subfigure[Centroid velocity response]{\includegraphics[width=0.495\columnwidth, trim= 10 5 85 30, clip]{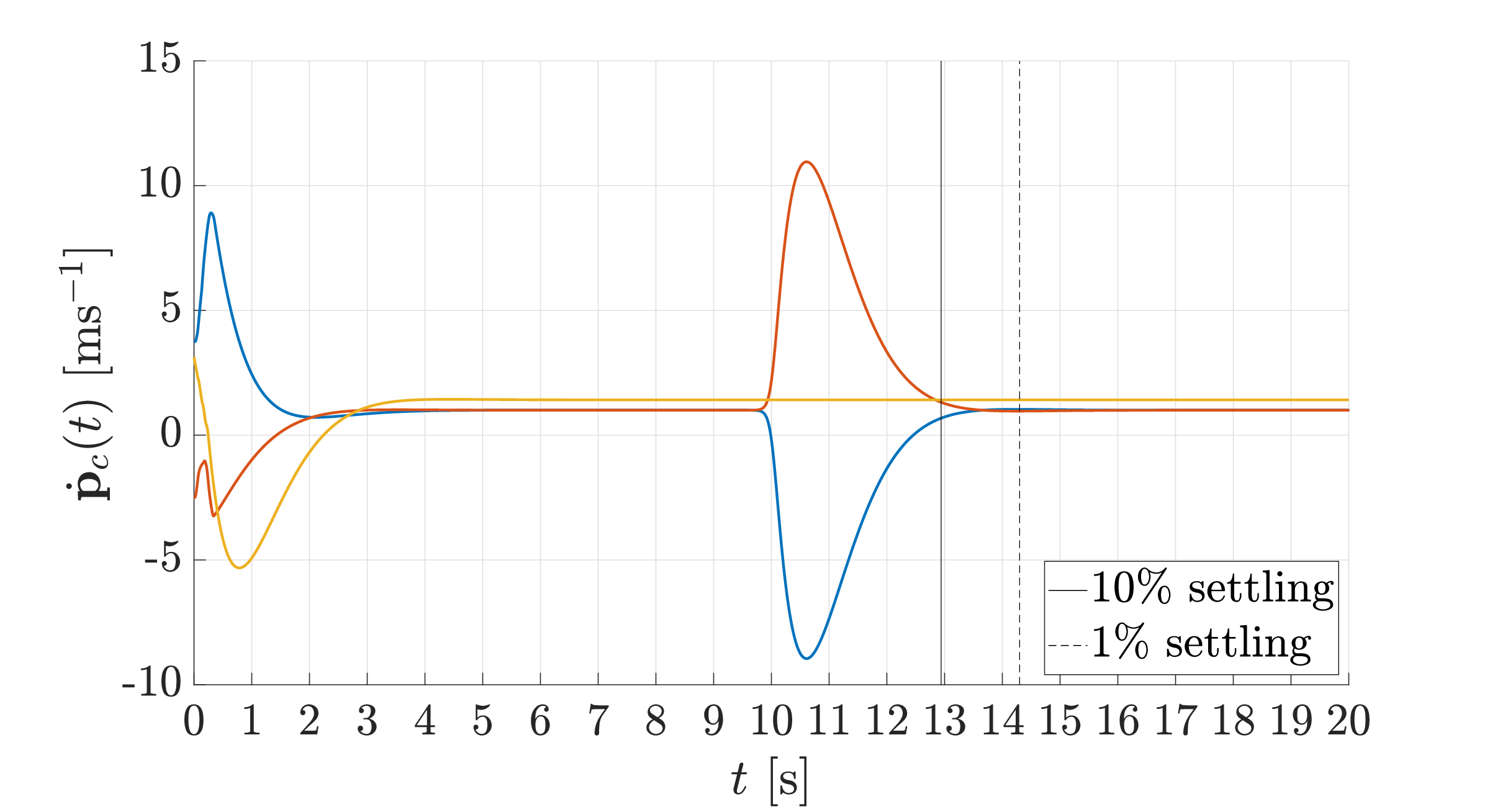}\label{fig:trackpdotc}}\vspace{-0cm}\\
	\caption{Distributed control law seen as a consensus protocol. (a)-(c)-(e): distance-based potential error functions \eqref{eq:potential} and their derivatives w.r.t. the corresponding current squared distance converging to $0$ before final time $T$; agents are guaranteed to reach the required geometrical shape. (b): relative velocities converging to $0$, ensuring the agents to follow the desired path with straight trajectories. (d)-(f): response of the centroid positions and velocities (first, second and third components in blue, orange and yellow respectively) converging to the given specifications.}
	\label{fig:consensusPRONTOdistr}		
\end{figure}

Fig. \eqref{fig:consensusPRONTOdistr} depicts an overview on the whole dynamics of the system governed by the online distributed feedback controller. Firstly, Figs. \eqref{fig:cons0sigma}, \eqref{fig:cons1sigma} and \eqref{fig:cons2sigma} exhibit the convergence to zero of the distance errors 
as $t \rightarrow T = 20~\si{\second}$. This shows the achievement of the formation positioning, i.e. the first-order cohesion/separation task. Secondly, in Fig. \ref{fig:reldistconsensus}, the relative velocities responsible for the second-order cohesion/separation tasks converge to zero. Moreover, Figs. \ref{fig:trackpc} and \ref{fig:trackpdotc} depict the response of positions and velocities of the centroid, showing that the tracking, namely the velocity alignment task, is fulfilled as required.
In each of these diagrams, it is worth to observe that the $1\%$ settling time instant computed through function $l^{tf}$ gives significant information about the convergence of the global state $\xx$ towards all the specifications. Indeed, by the rigidity properties of the desired formation, 
the entire dynamics controlled by law in \eqref{eq:distrOIFTsol} approximately attains a stable equilibrium after $13~\si{\second}$, for $t > t_{1\%}$, right after the effect of the chicane vanishes.
\section{Conclusions and future directions} \label{sec:conclusions}
Considering a second-order integrator MAS, we have introduced an extended version of the OIFT problem whose minimum-energy solution can be obtained by the numerical tool PRONTO (offline) and a distributed feedback control law (online). In order to fulfill the required final configurations for the formation to be driven, generic and versatile potential functions have been proposed. 
The reported simulations show the validity of the online distributed feedback control devised w.r.t. the trajectory and the energy profiles yielded by PRONTO, where the latter has been accounted as a reference for the former. 
Also, the flexibility of the decentralized paradigm is tested under the presence of communication constraints on the information sharing.


Several unresolved challenges remain that may be addressed in future research endeavors. As previously mentioned, further investigation will be dedicated to addressing the collision and obstacle avoidance issues, already explored in prior studies such as \cite{yang2019tracking, lee2017almost, zheng2023time, Huang2023Collision}. Another prospective avenue for this work involves extending its scope to scenarios where directed graphs model the information exchanged among agents, as demonstrated in \cite{mei2020robust, Babaz2022Directed}. Additionally, the exploration of the optimal time-varying formation tracking problem, as investigated in \cite{yu2022adaptive, zheng2023time}, poses an intriguing and challenging follow-up for the use of PRONTO.


\section*{Acknowledgments}
This work has been partly supported by the Grant ``Design Of Cooperative Energy-aware Aerial plaTforms for remote and contact-aware operations'' (DOCEAT) funded by the Italian MUR, n. 2020RTWES4, CUP n. E63C22000410001.

\bibliography{biblio_EJC}

\end{document}